\documentclass[aps,prb,twocolumn,superscriptaddress,floatfix,preprintnumbers,showpacs]{revtex4-2}
\usepackage[utf8]{inputenc}
\usepackage[english]{babel}
\usepackage{graphicx,graphics,amssymb,amsmath,bbm,appendix,ulem}
\usepackage{comment}
\usepackage{hyperref}
\usepackage[dvipsnames]{xcolor}

\newcommand{\unige}[0]{Department of Quantum Matter Physics, \'Ecole de Physique University of Geneva,
Quai Ernest-Ansermet 24, CH-1211 Geneva 4, Switzerland and}
\newcommand{\cea}[0]{Univ. Grenoble Alpes, CEA, Grenoble INP, IRIG-MEM-L\_Sim, Grenoble, France.}
\newcommand{\mf}[1]{{ \color{red} #1}}

\begin{document}
\title{Semi-classical theory of quantum stochastic resistors}
\author{Tony Jin}
\affiliation{\unige}
\author{Jo\~ao Ferreira}
\affiliation{\unige}
\author{Michel Bauer}
\affiliation{Universit\'e Paris-Saclay, CNRS, CEA, Institut de Physique Th\'eorique,
91191 Gif-sur-Yvette, France}
\affiliation{PSL Research University, CNRS, \'Ecole normale sup\'erieure, D\'epartement de math\'ematiques et applications, 75005 Paris, France}
\author{Michele Filippone}
\affiliation{\cea}
\author{Thierry Giamarchi}
\affiliation{\unige}

\begin{abstract}
We devise a semi-classical model to describe the transport properties of low-dimensional fermionic lattices under the influence of external quantum stochastic noise.
These systems behave as {\it quantum stochastic resistors}, where the bulk particle transport is diffusive and obeys the Ohm/Fick's law. Here, we extend previous exact studies beyond the one-dimensional limit to ladder geometries and explore different dephasing mechanisms that are relevant to different physical systems, from solid-state to cold atoms.  We find a non-trivial dependence of the conductance of these systems on the chemical potential of the reservoirs. We then introduce a semi-classical approach that is in good agreement with the exact numerical solution and provides an intuitive and simpler interpretation of transport in quantum stochastic resistors. Moreover, we find that the conductance of quantum ladders is insensitive to the coherence of the dephasing process along the direction transverse to transport, despite the fact that the system reaches different stationary states. We conclude by discussing the case of dissipative leads affected by dephasing, deriving the conditions for which they effectively behave as Markovian injectors of particles in the system. 

\noindent\begin{minipage}[t]{1\columnwidth}%
\global\long\def\ket#1{\left| #1\right\rangle }%

\global\long\def\bra#1{\left\langle #1 \right|}%

\global\long\def\kket#1{\left\Vert #1\right\rangle }%

\global\long\def\bbra#1{\left\langle #1\right\Vert }%

\global\long\def\braket#1#2{\left\langle #1\right. \left| #2 \right\rangle }%

\global\long\def\bbrakket#1#2{\left\langle #1\right. \left\Vert #2\right\rangle }%

\global\long\def\av#1{\left\langle #1 \right\rangle }%

\global\long\def\avb#1{\left\langle #1 \right\rangle _{\Omega_{\beta}}}%

\global\long\def\avi#1{\left\langle #1 \right\rangle _{\infty}}%

\global\long\def\nf#1{\left\Vert #1\right\Vert _{F}}%

\global\long\def\my{\text{\ensuremath{\mu_{y}}}}%

\global\long\def\mx{\mu_{x}}%

\global\long\def\tr{\text{tr}}%

\global\long\def\Tr{\text{Tr}}%

\global\long\def\haar{\text{Haar}}%

\global\long\def\pd{\partial}%

\global\long\def\im{\text{Im}}%

\global\long\def\re{\text{Re}}%

\global\long\def\sgn{\text{sgn}}%

\global\long\def\Det{\text{Det}}%

\global\long\def\abs#1{\left|#1\right|}%

\global\long\def\up{\uparrow}%

\global\long\def\down{\downarrow}%

\global\long\def\dag{\dagger}%

\global\long\def\cd{c^{\dagger}}%

\global\long\def\k{\mathbf{k}}%

\global\long\def\wks{\mathbf{\omega k}\sigma}%

\global\long\def\vc#1{\mathbf{#1}}%

\global\long\def\bs#1{\boldsymbol{#1}}%

\global\long\def\t#1{\text{#1}}%

\global\long\def\a{\mathbf{\alpha}}%

\global\long\def\b{\beta}%

\global\long\def\g{\gamma}%

\global\long\def\G{\mathbf{\Gamma}}%

\global\long\def\d{\mathbf{\delta}}%

\global\long\def\D{\mathbf{\Delta}}%

\global\long\def\e{\mathbf{\epsilon}}%

\global\long\def\l{\mathbf{\lambda}}%

\global\long\def\L{\mathbf{\Lambda}}%

\global\long\def\s{\sigma}%

\global\long\def\bs{\boldsymbol{\sigma}}%

\global\long\def\S{\Sigma}%

\global\long\def\r{\rho}%

\global\long\def\x{\chi}%

\global\long\def\H{\mathcal{H}}%


\global\long\def\w{\omega}%


\global\long\def\id{\mathbb{I}}%

\global\long\def\LL{\mathcal{L}}%

\global\long\def\ra{\rightarrow}%
\end{minipage}
\end{abstract}
\maketitle

\section{Introduction\label{sec:Introduction}}

Diffusion is the most common type of transport encountered in 
many-body systems, both in the classical and in the quantum world. In condensed matter setups, it is observed whenever the resistance of a metallic conductor is measured. The emergence of resistive behavior is commonly attributed to the diffusive propagation of charge carriers caused by scattering with disorder, impurities or particles of the same or different nature (electrons, holes, phonons, magnons \ldots)~\cite{akkermans_mesoscopic_2007}. Despite the clarity of these physical mechanisms, describing the emergence of diffusive transport from a full quantum perspective remains an open issue in theoretical physics~\citep{giamarchi_umklapp_1d,rosch_conservation_1d,lux2014hydrodynamic,medenjak2017diffusion,Gopalakrishnan2019,friedmanDiffusiveHydrodynamicsIntegrability2020,bertinifinite2021}. 

In recent years, the study of open quantum systems has opened new
exciting venues to understand the emergence of diffusion. The Markovian description of leads~\citep{Prosen2011,ProsenExactXXZ,KarevskiRepeatedinterac,KarevskiExactXXZ,ferreiraBallistictodiffusiveTransitionSpin2020a,JinFilipponeGiamarchi_GenericMarkovian,Znidaric__XXdeph,Bertini_2016}, losses~\citep{muller2021shape,rossini2020strong,Alba_Losses,Visuri_Losses}
or external time-dependent noises~\citep{Znidaric__dephasingXXZ,Znidaric__XXdeph,BastianelloDeNardisDeluca_GHDDeph,Eisler_CrossoverBallisticDiffusive,dolgirevNonGaussianCorrelationsImprinted2020a,Alba_EntanglementDephasing,BauerBernardJin_EquilibriumQSSEP,BernardJin_QSSEP,JinKrajenbrinkBernard_QKPZ,EsslerPiroli_Operatorfragmentation,BernardPiroli_QSSEPentanglement,BernardLeDoussal_StochasticCFT} has provided valuable numerical and analytic insight into
the problem. 
In this context, dephasing has been in the spotlight for being an analytically
tractable process of physical importance. It is capable of  describing the emergence of diffusion in quantum coherent systems~\cite{Znidaric__XXdeph,ProsenEssler_Mapping,BauerBernardJin_Stoqdissipative, TurkeshiSchiro_Dephmodel,TurkeshiSchiro_Dephmodel}, which behave as {\it quantum stochastic resistors}~\cite{Jin_Quantumresistors}.

\begin{figure}[ht!!]
\begin{centering}
\includegraphics[width=\columnwidth]{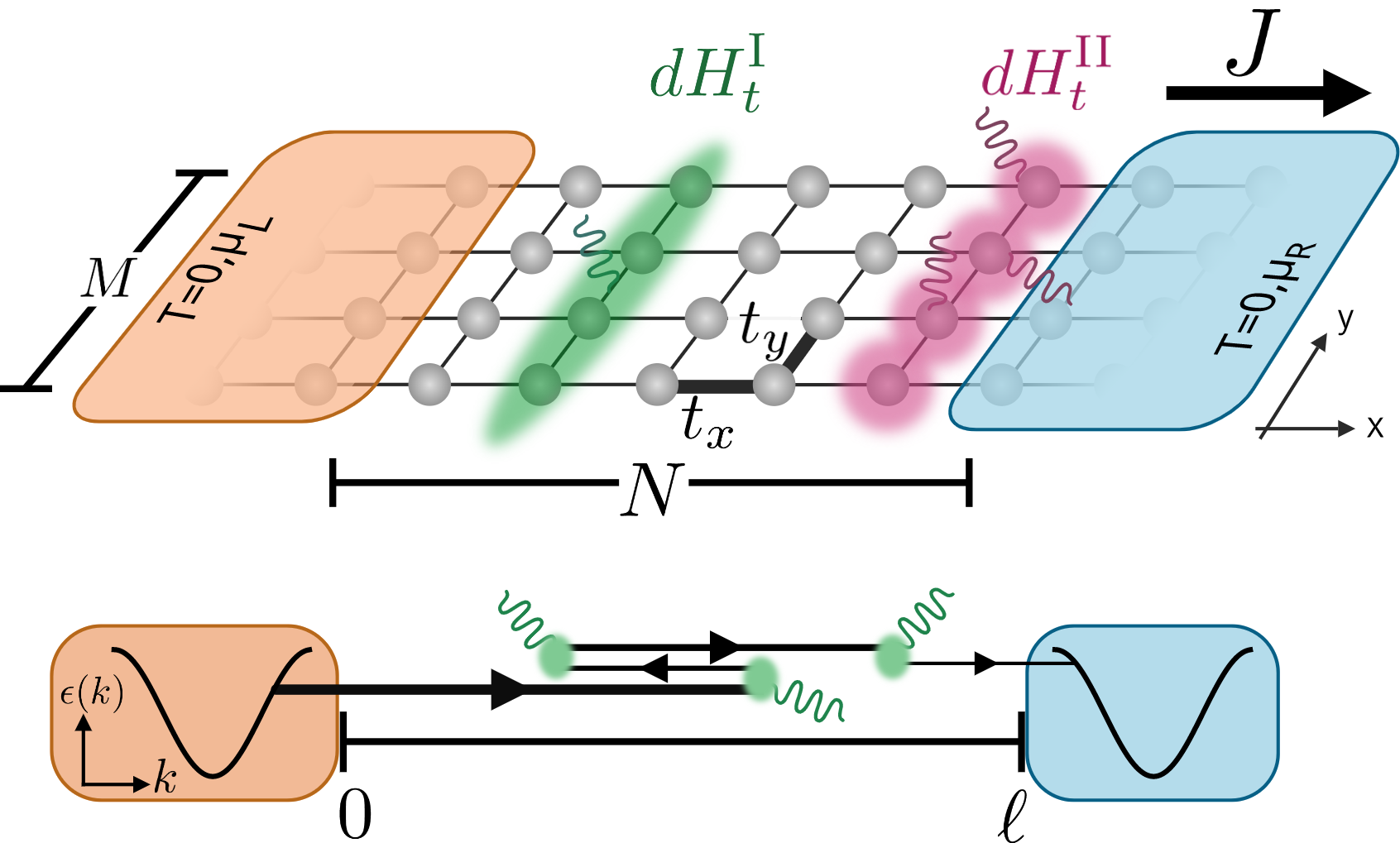}
\par\end{centering}
\caption{\label{fig:model}Top: Schematic representation of the system under study.
An $M$-leg square ladder is attached at the edges to two leads
prepared at the same temperature with distinct chemical potentials $\mu_{\rm{L},\rm{R}}$.
The bias $\protect\d\mu$ in the chemical potential drives a particle
current $J$, that can depend on the noise. Noises differ on the spatial correlation along the $y$-direction ranging from uniformly correlated, $dH_t^{\rm I}$, to uncorrelated, $dH_t^{\rm II}$.   Bottom:
Semi-classical interpretation of a 1D diffusive channel. A particle leaves a lead with a velocity determined by the band dispersion. A reset of
a particle's velocity occurs at random times until it escapes to one
of the leads. The distance between leads is $\ell=Na$ with $a$ the lattice spacing. 
}
\end{figure}

Despite these exact derivations of classical diffusive transport in the quantum realm, it remains an open question to which extent a classical description can account for the coherent transport properties and with which accuracy~\cite{denardisSuperdiffusionEmergentClassical2020,zuEmergentHydrodynamicsStrongly2021,wurtz_variational_2020}. If successful, a classical description could provide additional insight on transport phenomena outside the framework of open quantum systems.

Moreover, most of the studies mentioned above are restricted to one dimension, often exploiting integrable structures in some fine tuned cases~\cite{bertinifinite2021,ProsenEssler_Mapping,denardisDiffusionGeneralizedHydrodynamics2019,BastianelloDeNardisDeluca_GHDDeph}.
It is thus important to investigate the extension of exact solutions to higher dimensions and their richer behavior~\citep{steinigeweg2014scaling}. This understanding is also relevant to open new perspectives in the context of quantum matter simulators, where controlled dissipative dynamics is  under study in both bosonic~\cite{dogra2019dissipation,ferri_2021,rosamedina_2022} and fermionic systems~\citep{corman_2019,lebratquantized2019}.

In this work, we devise a semi-classical model which accurately describes the transport properties of low-dimensional quantum stochastic resistors. We focus on the quantum ladders geometries sketched in Fig.~\ref{fig:model}-top, where a current is driven by a difference of chemical potential $\delta \mu$ between thermal leads. 
The lattice is under the influence of dephasing processes and the working principle of the semi-classical model is illustrated in Fig.~\ref{fig:model}-bottom, in the one-dimensional limit. Semi-classically, dephasing is conceived as a stochastic reset of single particle velocities, which mimics a series of random quantum measurements of the particle position. 

To characterize the transport properties of the dephased ladders, we consider their conductance at a weak bias $\delta \mu$. We show that, in the presence of dephasing, the conductance is suppressed with the longitudinal extent of the system -- the number of sites $N$ in Fig.~\ref{fig:model}-top -- revealing the emergence of bulk resistivity. We also observe that dephasing triggers a non-trivial dependence of the conductance on the chemical potential $\mu$ of the reservoirs. In particular, the conductance vanishes when the chemical potential approaches the band edges, reflecting a suppression with the velocity of particles injected by the reservoirs. This dependence is absent in the ballistic case and is particularly intriguing as it is  also absent in the bulk diffusion constant of the system. We show then how the semi-classical model is able to accurately reproduce the emergent $\mu$-dependence of the  conductance, providing at the same time a simple physical picture connecting boundary and bulk diffusive effects. 

We then extend these considerations to ladder systems. The presence of an additional degree of freedom along the $y$-direction, transverse to the current flow along $x$, allows different dephasing processes. These processes can be either coherent or incoherent along the $y$-direction, see Fig.~\ref{fig:model}-top. The coherent case is for instance relevant to cold atom systems with a  synthetic $y$-dimension~\cite{Mancini1510,genkina2019imaging,chalopin_probing_2020,zhou2022observation}. Even though these different noises drive the system towards totally different stationary states, we find that they carry exactly the same current. We explain this remarkable coincidence as a manifestation of the fact that the correlations of these different noises obey identical isotropy conditions, that we derive and discuss in detail.

Finally, we study the case where the leads themselves are dissipative and affected by dephasing. In the limit where the dephasing rate $\gamma$ is large compared to other characteristic energy scales of the system, we show that the action of the leads can be mapped to that of a Lindblad boundary injection and extraction, which is controlled by the density of particle at the system edges.

This paper is structured as follows. In Section~\ref{sec:Model}, we discuss the Keldysh approach for the exact self-consistent derivation of currents in quantum stochastic ladder resistors. Section~\ref{sec:1D-diffusive_channel} introduces the semi-classical approach and illustrates its ability to reproduce exact results. Section~\ref{sec:Dephasingladder} discusses the extension to ladders and Section~\ref{sec:Out-of-equilibrium-leads} the effective description of dissipative leads affected by dephasing. Section~\ref{sec:Conclusion} discusses  results and conclusions.

\section{Model and Methods\label{sec:Model}}


We study the transport properties of spinless fermions on the discrete square lattice geometry sketched in Fig.~\ref{fig:model}-top. We consider an infinite lattice along the longitudinal direction
($x$-axis), with $M$ sites in the transverse direction ($y$-axis).
The corresponding Hamiltonian reads 
\begin{equation}\label{eq:system}
H=-\sum_{j,m}\left[t_{x}c_{j+1,m}^{\dagger}c_{j,m}+t_{y}c_{j,m+1}^{\dagger}c_{j,m}+{\rm h.c}\right]\,,
\end{equation}
where the sum over $j$ runs between $\pm\infty$, while $m=1,\dots,M$. The  operators $c_{j,m}$ annihilate fermions on site $(j,m)$ and $t_{x/y}$ control the hopping amplitude along the $x/y$ directions. We  further divide the sum over the longitudinal direction into three regions: the system (S) for $j\in[1,N]$, the left (L) lead for $j<1$ and the right (R) lead for $j>N$, see Fig.~\ref{fig:model}-top. It is useful to introduce the basis diagonalizing the transverse hopping term in Eq.~\eqref{eq:system}, given by the unitary transformation $a_{j,p}=\sum_{m=1}^{M}\sqrt{\frac{2}{M+1}}\sin\left(\frac{\pi mp}{M+1}\right)c_{j,m}$. This transformation uncouples the $M$ transverse modes and the corresponding Hamiltonian reads
\begin{align}\label{eq:systemdiag}
H= \sum_{j,p}\left[-t_{x}\left(a_{j+1,p}^{\dagger}a_{j,p}+{\rm h.c}\right)+\epsilon_pa_{j,p}^{\dagger}a_{j,p}\right]\,,
\end{align}
with $\epsilon_p=-2t_{y}\cos\left(p\pi/(M+1)\right)$ and $p\in[1,M]$. If the system is translational invariant along the $x$-direction, the transverse modes have non-degenerate dispersion relations $\e_{p,k}=-2t_{x}\cos\left( k\right)+\epsilon_p$, with $k\in[-\pi,\pi]$ the quasi-momentum in the first Brillouin zone, see sketches in Fig.~\ref{fig:2-leg-ex} for an illustration in the $M=2$ case. We reserve the indexes $j,m$ for the physical sites in the $x$ and $y$ direction, and the indexes
$k,p$ label respectively longitudinal quasi-momenta and transverse modes. 

In addition to the coherent Hamiltonian dynamics, we introduce a noise term modelled by a quantum stochastic Hamiltonian (QSH) that leads to various dephasing mechanisms that we are going to detail. The QSH is defined by the infinitesimal generator $dH_{t}$ such that the total unitary operator $U(t)$ is evolved as
\begin{equation}
    U(t+dt)=e^{-i(Hdt+dH_{t})}U(t)\,.
\end{equation}
In this work, we are interested in QSHs which conserve the total particle number and lead to dephasing. They are described by 
\begin{equation}\label{eq:deph_hamiltonian}
dH_{t}=\sqrt{2\gamma}\sum_{j,p,p'}a_{j,p}^{\dagger}a_{j,p'}dW_{t}^{j,p,p'}\,,\end{equation}
where $\gamma$ controls the overall dephasing rate and the $dW_t$ are increments of stochastic processes defined within the It\=o prescription~\cite{Oksendal_book} with zero mean and covariance
\begin{equation}\label{eq:Ito}
\begin{aligned}
dW_{t}^{j_{1},p_{1},p'_{1}}dW_{t'}^{j_{2},p_{2},p'_{2}}&=\delta_{j_{1},j_{2}}C_{p_{1},p'_{1},p_{2},p'_{2}}dt &\text{for } t=t',\\
dW_{t}^{j_{1},p_{1},p'_{1}}dW_{t'}^{j_{2},p_{2},p'_{2}}&=0 &\text{for } t\neq t'.
\end{aligned}
\end{equation}

By construction, the noise is thus uncorrelated in time and in the longitudinal $x$-direction, $j$ index, but not necessarily on the transverse $y$-direction, $p$ index. Correlations of the noise in the $y$-direction are taken into account by the function $C$, which can be adapted to describe different physical scenarios, as we are going to illustrate in the context of ladder geometries in Section~\ref{sec:Dephasingladder}. Since the $dW_{t}$ commute with one another, we have $C_{p_{1},p'_{1},p_{2},p'_{2}}=C_{p_{2},p'_{2},p_{1},p'_{1}}$. Hermiticity also imposes that  $dW_{t}^{j,p,p'}=(dW_{t}^{j,p',p})^*$.
Qualitatively speaking, each term in the sum of Eq.~\eqref{eq:deph_hamiltonian} describes transitions from a state indexed by $p'$ to a state indexed by $p$ with a random complex amplitude given by $dW_t^{j,p,p'}$. Since the $C$s are arbitrary, Eq.~\eqref{eq:deph_hamiltonian} constitutes the most general way of writing noisy quadratic jump processes between different transverse propagation modes. In one-dimension, discussed in Section~\ref{sec:1D-diffusive_channel}, Eq.~\eqref{eq:deph_hamiltonian} reduces to an on-site stochastic fluctuation of potential, leading to standard dephasing, see also Eqs.~(\ref{eq:deph1D}-\ref{eq:1DdephasingLindblad}). In Section~\ref{sec:Dephasingladder}, we will specify different noise-correlations on ladders and discuss their implication on transport.

The mean evolution generated by the stochastic Hamiltonian~\eqref{eq:deph_hamiltonian}, with the prescription~\eqref{eq:Ito}, is described by the Lindblad generator acting on the reduced density matrix  of the system $\rho$ 
\begin{multline}\label{eq:lindblad}
{\cal L}(\rho)=  \gamma\sum_{j,p_{1},p_2,p_{1}',p_2'}C_{p_{1},p'_{1},p_{2},p'_{2}}(2a_{j,p_{1}}^{\dagger}a_{j,p'_{1}}\rho a_{j,p_{2}}^{\dagger}a_{j,p'_{2}} \\
  -\{a_{j,p_{2}}^{\dagger}a_{j,p'_{2}}a_{j,p_{1}}^{\dagger}a_{j,p'_{1}},\rho\})\,,
\end{multline}
where $\{,\}$ denotes anticommutation.

\subsection{Keldysh approach and exact self-consistent solution of transport in quantum stochastic resistors}
As we will be dealing with systems under the effect of dephasing noise and biased leads, the dynamics of the system is intrinsically out of equilibrium. The natural language to describe these systems is  the Keldysh formalism~\cite{Kamenev2011}, detailed in App.~\ref{App subsec: Green'sfunction}. 
The central objects of the theory are  the retarded ($R$), advanced ($ A$) and Keldysh ($K$) components of the single-particle Green's functions $\mathcal{G}^{R/A/K}$. They are defined in time representation as  $\mathcal{G}^{R}_{j,m;i,n}(t-t')=-i\theta(t-t')\langle\{c_{j,m}(t),c^\dagger_{i,n}(t')\}\rangle$, $\mathcal{G}^{ A}_{j,m;i,n}(t-t')=[\mathcal{G}^{ R}_{i,n;j,m}(t'-t)]^*$ and $\mathcal{G}^{ K}_{j,m;i,n}(t-t')=-i\langle[c_{j,m}(t),c^\dagger_{i,n}(t')]\rangle$
~\footnote{The Green's functions depend on the time differences $t-t'$, instead of separate times $t,t'$, as we consider situations where both the Hamiltonian~\eqref{eq:system} and Lindblad generator~\eqref{eq:lindblad} do not depend explicitly on time.}. By adopting the notation by Larkin and Ovchinnikov~\cite{Larkin_vortices_supra}, these three components are collected in a unique matrix, which obeys the Dyson equation
\begin{align}\label{eq:Dyson}
\boldsymbol{\mathcal{G}}&=\begin{pmatrix}\mathcal{G}^{R} & \mathcal{G}^{K}\\
0 & \mathcal{G}^{A}
\end{pmatrix}\,, & \boldsymbol{\mathcal{G}}^{-1}&=\boldsymbol{g}^{-1}-\boldsymbol{\Sigma}\,,
\end{align}
where $\boldsymbol g$ corresponds to the Green's function of the system disconnected from the leads and unaffected by noise. The matrix $\boldsymbol \Sigma$ corresponds to the self-energy, which has the same matrix structure as $\boldsymbol{ \mathcal G}$.

In the path integral formalism, the fermionic degrees of freedom of the leads can be integrated out. Their integration gives a contribution to the self-energy of the system $\boldsymbol{\S}_{\text{L/R}}$, which has non-zero components only at the system edges $j=1,N$. The general procedure of this integration is detailed in  App.~\ref{App subsec: Green'sfunction}. To give a more explicit idea of the result of this procedure, we report here the result for the simplest one-dimensional case ($M=1$). The edge contributions then read
\begin{equation}\label{eq:self-leads}
\begin{aligned} & \Sigma_{{\rm L},i,j}^{R/A}=t_{x}^{2}g_{0,0}^{R/A}\delta_{i,j}\delta_{i,1},\\
 & \Sigma_{{\rm R},i,j}^{R/A}=t_{x}^{2}g_{N+1,N+1}^{R/A}\delta_{i,j}\delta_{i,N},\\
 & \Sigma_{{\rm L},i,j}^{K}=2it_{x}^{2}F_{{\rm L}}\Im(g_{1,1}^{R})\delta_{i,j}\delta_{i,1},\\
 & \Sigma_{{\rm R},i,j}^{K}=2it_{x}^{2}F_{{\rm R}}\Im(g_{N,N}^{R})\delta_{i,j}\delta_{i,N},
\end{aligned}
\end{equation}
where $\Im(\cdot)$ gives the imaginary part. The retarded and advanced components of the self-energy are renormalized by the corresponding reservoir Green functions,  which are calculated at the site closest to the system. See Eq.~\eqref{eq:grl11} for the explicit expression of $g^{R/A}_{0,0}$ and $g^{R/A}_{N+1,N+1}$ in the case of leads identical to the system. The Keldysh components describe the tendency of the edges of the system to equilibrate to the attached reservoirs. The functions $F_{\rm L,R}$ describe the state of the leads, and the self-energies obey a local equilibrium fluctuation-dissipation relation~\cite{Kamenev2011}. In the absence of noise, the leads are considered in thermal equilibrium with a well-defined chemical potential $\mu_{\rm L,R}$ and shared temperature $T$. In frequency representation, this situation is described by $F_{\rm L,R}(\omega)=\tanh[(\omega-\mu_{\rm L,R})/2T]$. The fact that the system is out of equilibrium can be read in Eq.~\eqref{eq:self-leads} via the fact that different functions $F$ affect the self-energy of the system at its borders.  We will also consider the case of diffusive leads affected by dephasing in Section~\ref{sec:Out-of-equilibrium-leads}.

The Keldysh formulation of the problem is advantageous because it allows to deal exactly with the dephasing dynamics caused by the presence of the noise described by Eq.~\eqref{eq:deph_hamiltonian}. Despite the quartic nature
of Eq.~\eqref{eq:lindblad}, the stochastic formulation of the dephasing~\eqref{eq:deph_hamiltonian} allows for a closed exact solution of the
self-energy \citep{dolgirevNonGaussianCorrelationsImprinted2020a,TurkeshiSchiro_Dephmodel,Jin_Quantumresistors}. Indeed, the latter can be expressed in terms of the Green's function via the relation
\begin{multline} \label{eq:self-energy}
  \boldsymbol{\Sigma}_\gamma(t,t')_{(j,p_{1}),(j',p'_{2})}= \\
  \gamma\delta(t-t')\delta_{j,j'}\sum_{p'_{1}p_{2}}C_{p_{1},p'_{1},p_{2},p'_{2}}\boldsymbol{{\cal G}}_{(j,p'_{1})(j,p_{2})}(t,t)\,,
\end{multline}
which, inserted in Eq.~\eqref{eq:Dyson}, has to be solved self-consistently. To summarize, we derive an explicit expression of the self-energies in the Dyson equation~\eqref{eq:Dyson}, which reads
\begin{equation}\label{eq:Dyson1}
\boldsymbol{\mathcal{G}}^{-1}=\boldsymbol{g}^{-1}-\boldsymbol{\S}_{\rm{L}}-\boldsymbol{\S}_{\rm{R}}-\boldsymbol{\Sigma}_{\g}\,,
\end{equation}
where the expression of $\boldsymbol{\mathcal{G}}$ is obtained numerically.

Equipped with the formal expression of the single-particle Green's functions, we can directly and exactly inspect the transport properties of quantum systems under the influence of dephasing noise. By imposing a finite bias, $\mu_{{\rm \text{L/R}}}=\mu\pm\frac{\delta\mu}{2}$,
between the right and left leads, a uniform longitudinal 
current $J$ flows through the system. By construction, the noise \eqref{eq:deph_hamiltonian} preserves the total density $n_{j,{\rm tot}}=\sum_{p}a_{j,p}^{\dagger}a_{j,p}$
at a fixed position $j$ on the $x$-axis, i.e $[dH_{t},n_{j,{\rm tot}}]=0.$
Thus, the definition of the total longitudinal current
operator is
unchanged by the noise term, and the current can be evaluated at any site $j$, namely  
\begin{align}\label{eq:current}
J&=it_{x}\sum_{m=1}^{M}\av{c_{j+1,m}^{\dagger}c_{j,m}-c_{j,m}^{\dagger}c_{j+1,m}} \nonumber\\
&
=\frac{t_{x}}{2}\sum_{m=1}^{M}\int\frac{d\omega}{2\pi}\left(\mathcal{G}_{j,m;j+1,m}^{K}-\mathcal{G}_{j+1,m;j,m}^{K}\right)(\omega)\,.
\end{align}
In the following, we will explicitly derive this expression from the exact self-consistent solution of Dyson's equation~\eqref{eq:self-energy}. Additionally, we rely on the linear expansion of Eq.~\eqref{eq:current} in the chemical potential difference $\delta\mu$ to study the conductance of the system, which is defined as
\begin{equation}\label{eq:conductance}
G=\lim_{\delta\mu \rightarrow 0}2\pi\frac{J}{\d\mu}\,.
\end{equation}
Notice that we rescaled the conductance by $2\pi$ in order to have the quantum of conductance equal to $1$ and adopt the convention $e=k_B=\hbar=1$. In the following sections, we devise a semi-classical model which can capture the results from Eqs.~\eqref{eq:Dyson1}, \eqref{eq:current} and~\eqref{eq:conductance} in the presence of dephasing. In particular, we will inspect the conductance dependence on the chemical potential of the leads $\mu$.


\section{Conductance of a 1D quantum stochastic resistor \label{sec:1D-diffusive_channel}}
In this section, we focus on a strictly one-dimensional geometry to showcase the effectiveness of the semi-classical approach in describing the emergent diffusive transport properties of quantum stochastic resistors. 

\subsection{Exact derivation}\label{sec:exact}
We begin by deriving the dependence of the conductance $G$ on the chemical potential of the leads $\mu$, via the exact solution of a single chain subjected
to on-site dephasing noise. In one-dimension, the noise term in Eq.~\eqref{eq:deph_hamiltonian} reduces to
\begin{equation}\label{eq:deph1D}
dH_{t}=\sqrt{2\gamma}\sum_{j}c^{\dag}_j c_j dW_{t}^{j}\,,
\end{equation}
with the corresponding Lindblad operator
\begin{equation}
{\cal L}(\rho)=\gamma\sum_{j}\left(2n_{j}\rho n_{j}-\{n_{j},\rho\}\right).\label{eq:1DdephasingLindblad}
\end{equation}

For a QSH described by Eq.~\eqref{eq:deph_hamiltonian}, the conductance of the system can be derived from a generalized expression of Meir-Wingreen's formula~\cite{JinFilipponeGiamarchi_GenericMarkovian,MeirWingreenformula}
\begin{equation}
G_{\g}(\mu)=\int d\w  \frac{\mathcal{T}_{\g}(\w)}{4T \cosh^2\left(\frac{\w-\mu}{2T}\right)}\,.\label{eq:LB}
\end{equation}
This expression  for the conductance reproduces Landauer-Büttiker's formula, valid for non-interacting ballistic systems~\citep{LandauerFormula}. As such, $\mathcal{T}_{\g}(\w)$ is interpreted as the transmittance of the channel at energy $\w$ for a fixed $\gamma$, a quantity independent of the temperature $T$ and chemical potential $\mu$ of the leads. An explicit expression of $\mathcal T_{\g}(\omega)$ was computed in Ref.~\citep{Jin_Quantumresistors} in similar settings.  We stress that the extension of Landauer-Büttiker's formula~\eqref{eq:LB} to dephased systems is highly non-trivial, given the fact that dephasing triggers inelastic scattering events in the conducting region.  

\begin{figure}[t]
\begin{centering}
\includegraphics[width=0.95\columnwidth]{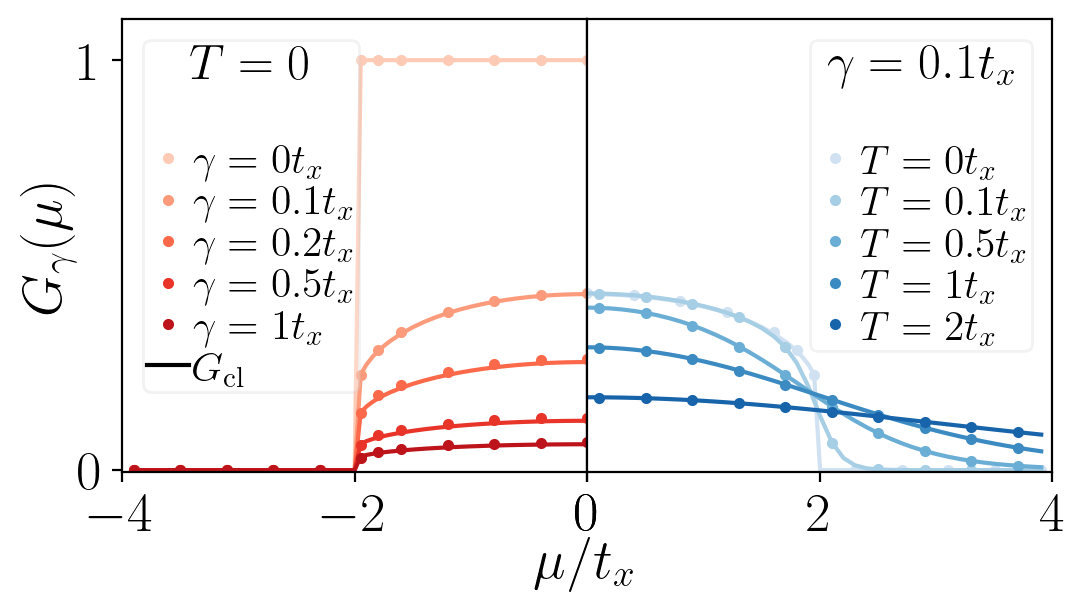}\linebreak{}
\includegraphics[width=0.95\columnwidth]{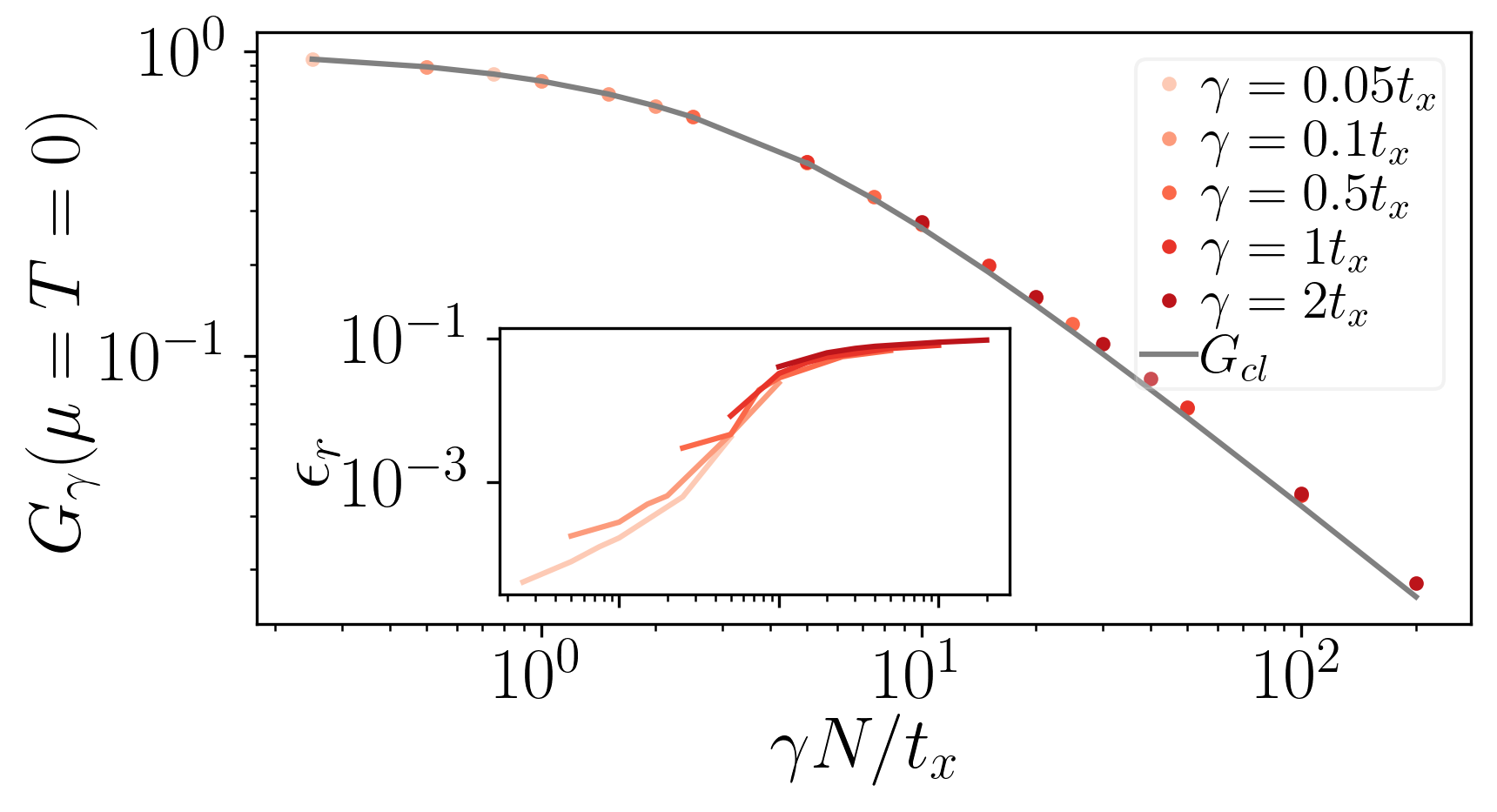}
\par\end{centering}
\caption{\label{fig:diffusive}\textcolor{black}{Top: Conductance as a function
of chemical potential for increasing dephasing rates $\protect\g$ at $T=0$ (left) and increasing temperature (right)
at a fixed system size $N=50$. The dots are derived relying on the exact quantum calculation (Section~\ref{sec:exact}), while the dashed lines correspond to  the semi-classical approximation (Section~\ref{subsec:Semi-classical-1D}). Bottom: Scaling of the conductance with the parameter $\protect\g N$ when $T=0,\mu =0$. Inset: relative error of the semi-classical conductance at $\mu=0$, $\e_{r}=|G_{cl}-G_{\g}|/G_{\g}$.}}
\end{figure}

If we consider leads which are identical to the system, see Eq.~\eqref{eq:system}, no reflection occurs at the interface and $\mathcal{T}_{\g=0}(\w)=1$ for $\w\in[-2t_{x},2t_{x}]$ and $0$ elsewhere.
At zero temperature, this implies the usual quantized conductance $G=1$ when the chemical potential of the leads lies within the dispersion relation of the reservoirs, $\mu\in[-2t_x,2t_x]$~\cite{datta_electronic_1997,Lesovik_2011,nazarov_quantum_2009,akkermans_mesoscopic_2007}, see Fig.~\ref{fig:diffusive}.

The presence of any finite dephasing rate leads to diffusive
transport in the thermodynamic limit~\citep{Znidaric__XXdeph,Znidaric_dephasing,BauerBernardJin_Stoqdissipative,TurkeshiSchiro_Dephmodel,Jin_Quantumresistors}.  In these studies, it was shown that the bulk transport properties are described by Fick's law
\begin{equation}\label{eq:fick}
    J=-D\nabla n\,,
\end{equation}
where $D$ is the diffusion constant and $\nabla n$ the particle density gradient along the chain. In particular, for fixed boundary conditions, Fick's law implies the $1/N$ suppression of the current with the system size and \begin{equation}\label{eq:D}
    D=\frac{2t_x^2}{\gamma}\,.
\end{equation}
This suppression reveals the emergence of a resistive behavior, compatible with Ohm's law. This relation holds in the bulk \textit{regardless} of the average chemical potential $\mu$ and temperature $T$ of the biased leads. This fact can be understood as follows: at equilibrium, the effect of the noise term is to drive the system towards an infinite temperature state with a fixed number of particle \citep{CaiZi_AlgebraicvsExponentialDissipativesystems}. Here the situation is more intricate since we are out-of-equilibrium. Nevertheless, we show numerically in  App.~\ref{App subsec: Heat} that, deep in the bulk, there exist a well-defined notion of \textit{local equilibrium}, where the system does reach an infinite temperature state. Thus, in the bulk, the information about the energy scales of the leads is erased, and one expects that bulk transport properties, such as the diffusion constant,  will be independent of the temperature and the chemical potentials of the boundaries. This point will be further emphasized in Sec.~\ref{sec:Dephasingladder}.

In contrast to the diffusion constant $D$, the conductance strongly depends on the temperature and chemical potential of the attached leads, see Fig.~\ref{fig:diffusive}-top~\footnote{Despite the possibility to rely on Eq.~\eqref{eq:LB} to calculate the conductance for $\gamma>0$, we found more practical to perform the direct numerical calculation of the current as expressed in Eq.~\eqref{eq:current} directly in the linear regime to derive the conductance~\eqref{eq:conductance}} . For a finite dephasing rate $\g$, $G_{\g}$ develops a clear dome-like dependence on the chemical potential~\footnote{The different scalings of $D$ and $G$ with $\mu,T$ indicate that the contact resistance between bulk and leads is extensive with the system size. Fig.\ref{fig:heat} and previous works~\citep{TurkeshiSchiro_Dephmodel} suggest that thermalization only occurs very deep in the bulk, supporting this hypothesis.} This shape persists even in the diffusive regime $ N\gg1/\gamma$ where the conductance vanishes as $1/N$, see Fig.~\ref{fig:diffusive}-bottom and Fig.~\ref{fig:dome} in  App.~\ref{App subsec: Conductance}. At $T=0$, the dome is restricted to energies within the bandwidth  $[-2t_x,2t_x]$ and the differential conductance $\partial_\mu G$ diverges whenever the chemical potential touches the edges of the band, even when $\g>0$. This  behavior is reminiscent of the ``staircase'' behavior of the conductance for non-interacting systems and $\gamma=0$. The main difference is that for $\gamma>0$ the conductance $G$ is not quantized and acquires a $\mu$ dependence in the $[-2t_x,2t_x]$ interval. As expected, increasing the temperature of the leads smears the dependence of the conductance with respect to the chemical potential, as illustrated in Fig.~\ref{fig:diffusive}-top.

The emergence of a dome-like dependence of the conductance ultimately originates from its non-local character which strongly depends on the geometry of the system, in this case, the connection to leads. Nevertheless, its qualitative shape prompts a physical explanation which is difficult to extract from the exact, numerical solution.

In the next section, we show that a semi-classical model allows to build an intuitive physical explanation of the dependence of $G_\gamma$ on the chemical potential and to connect it with the bulk behavior of transport.

\subsection{Semi-classical approach\label{subsec:Semi-classical-1D}}

The Lindblad operator~\eqref{eq:1DdephasingLindblad} can actually describe the average evolution of a system under different stochastic processes, which differ from the stochastic fluctuations of potential considered in Eq.~\eqref{eq:deph1D}. Indeed, the most natural way to devise a semi-classical description of the Lindblad dynamics of Eq.~\eqref{eq:1DdephasingLindblad} is to ``unravel'' it to a projective measurement process, where the densities at each site are measured independently with rate $\gamma$~\cite{Dalibard_Unraveling,Belavkin_1990}. Notice that for single realizations of the stochastic process, the projective dynamics fundamentally differs from the quantum stochastic dynamics described by Eq.~\eqref{eq:deph1D}. For instance, a density measurement on site $j$ would project the system in a state with 1 or 0 particles on that site in a non-unitary fashion. On the contrary, the random potential fluctuations described by Eq.~\eqref{eq:deph1D} are always unitary at the level of a single realization. Nevertheless,  the projective and QSH dynamics coincide in average and are described by the same effective Lindblad operator~\eqref{eq:1DdephasingLindblad}. 

In the projective case, at each time step $\Delta t$, a measurement at site $j$ occurs with probability $\gamma\Delta t$.
After a measurement, depending on whether the local particle number is measured to be zero or one, the density matrix is updated as follows:
\begin{equation}
\begin{aligned}
\rho&\to\rho_0=\frac{(1-n_{j})\rho(1-n_{j})}{{\rm Tr}[\rho(1-n_{j})]}\,,&
\rho_1=\frac{n_{j}\rho n_{j}}{{\rm Tr}[\rho n_{j}]}\,,
\end{aligned}
\end{equation}
with respective probabilities
\begin{align}
P_{\rho_0}&={\rm Tr}[\rho(1-n_{j})]\,,&    P_{\rho_1}&={\rm Tr}[\rho n_{j}]\,.
\end{align}
Averaging over the possible outcomes for a small time step $dt$ yields the average evolution of the density matrix $d\rho_{t}=\rho_{t+dt}-\rho_t$:
\begin{align}
d\rho_{t}= & \gamma dt\sum_{j}\big(2n_{j}\rho_{t}n_{j}-\{n_{j},\rho\}\big)\,,
\end{align}
which is equivalent to the Lindblad evolution described by Eq.~\eqref{eq:1DdephasingLindblad}.

This alternative point of view is the natural one to devise a semi-classical description of transport in systems affected by dephasing. If we consider a single-particle traveling through the chain, the effect of a measurement is to localize it
at a given site $j$. When the particle is localized, it is in a superposition of all possible momentum states.  

We thus propose the analogous classical model in the continuum limit: consider a single particle of initial velocity $v_{0}(\w)$ coming from the left lead into the system of length $\ell=Na$, where $a$ is the lattice spacing. Its velocity is set by its energy $\omega$, $v_{0}(\w)=\partial\epsilon_k/\partial k\vert_{\w}$,
where $\epsilon_k$ is the dispersion relation of the lead,
see Fig.~\ref{fig:model}-bottom. At a random time $t$, determined
by the Poissonian probability distribution $p(t)=\gamma e^{-\gamma t}$,
its velocity is reinitialized by drawing a momentum $k$ sampled from a uniform probability distribution on the interval $[-\pi,\pi]$. For a  dispersion relation $\epsilon_k=-2t_x\cos(k)$,
the probability distribution of the velocity $v$ reads
\begin{equation}\label{eq:pv}
p(v)=\frac{1}{2\pi t_{x}\sqrt{1-\left(\frac{v}{2t_{x}}\right)^{2}}},\quad v\in[-2t_{x},2t_{x}].
\end{equation}
Once the velocity has been reset, the process is restarted. Whenever
the particle reaches one boundary located at $x=0$ or $x=\ell$,
it exits the system. The problem of computing the semi-classical transmittance $\cal{T}_{{\rm cl}}$ can be reduced to compute the probability of exiting the system by
touching the right boundary. Note that this problem differs from a usual random walk, as in this case the length of the steps are not uniform in time. 

Once a measurement occurs, the velocity of a particle injected by a reservoir gets totally randomized according to the probability distribution~\eqref{eq:pv}. Thus the object of interest becomes the probability $P(x)$ of exiting the system once a given measurement has taken place at some position  $x\in[0,\ell]$. The first measurement takes place at position $x$ and time $t=x/v_0(\omega)$ with Poissonian probability distribution $\gamma e^{-\gamma t}$.  Thus, the semi-classical transmittance $\mathcal T_{\rm{cl}}$,  for a particle injected from the left lead with velocity $v_0(\omega)$, is given by 
\begin{equation}\label{eq:tcl}
\mathcal{T}_{{\rm cl}}(\omega)=
\int_0^{\infty} P(x)\frac{\gamma}{v_{0}(\w)}e^{-\gamma\frac{x}{v_{0}(\w)}}dx\,.
\end{equation}
We recall that, because of the specific dispersion of the leads under consideration, $\mathcal{T}_{\rm cl}(|\omega|>2t_x)=0$. 

It remains to determine  $P(x)$. It is useful to introduce the probability $P_v(x)$ for a particle to exit on the right when
it starts at $x$ with velocity $v$. The probability $P(x)$ is thus the integral of this probability over all possible velocities,  $P(x)=\int dv \,p(v)P_v(x).$ As we assume that no measurement process occurs in the leads, $P(x)$ has to fulfill the boundary conditions
\begin{align}
P(x<0)&=0\,,&P(x> \ell)&=1\,.
\end{align}
In the system, where the measurement processes occur, $P_v(x)$ is expressed in the closed form 
\begin{multline}\label{eq:Proba}
P_v(x)= \theta(v)\left[e^{-\gamma\frac{\ell-x}{v}}+\int_{0}^{\frac{\ell-x}{v}}dt\gamma e^{-\gamma t}P(x+vt)\right]\\
  +\theta(-v)\int_{0}^{-\frac{x}{v}}dt\,\gamma e^{-\gamma t}P(x+vt)\,, 
\end{multline}
where $\theta(v)$ is the usual Heaviside step function. The first term corresponds to the probability that the particle goes through the system without the occurrence of any measurement. 
The second term is the probability that a right mover resets at time $t$ multiplied by the probability to exit if the particle starts again from this position. The last term corresponds to the same process but for a left mover. By
integrating over the distribution of velocities~\eqref{eq:pv}, we get an implicit equation for $P(x)$ for $x\in[0,\ell]$:
\begin{multline}\label{eq:pximplicit}
P(x)=  \varphi(\ell-x)-\int_{0}^{\ell-x}dy\,\varphi'(y)\,P(x+y)\\-\int_{0}^{x}dy\,\varphi'(y)\,P(x-y)\,, 
\end{multline}
where we have introduced the function
\begin{align}
\varphi(y) =\int_{0}^{\infty}dv\,p(v)\,e^{-\gamma\frac{y}{v}} =\int_{0}^{1}dx\,\frac{e^{-\frac{\gamma y}{2t_xx}}}{\pi\sqrt{1-x^{2}}}\,,
\end{align}
and $\varphi'(y)=\partial \varphi/\partial y$. From Eq.~\eqref{eq:pximplicit}, the probability $P(x)$ can be in principle derived iteratively in the number of measurement-induced resets of velocity. This solution would consist in writing 
\begin{equation}\label{eq:recursion}
P(x)=\sum_{n=0}^{\infty}P_{n}(x)\,,
\end{equation}
where $P_{n}(x)$ is the probability of exiting on the left after
$n$ resets starting from $x$. This leads to
\begin{align}
P_{0}(x)=&\varphi(\ell - x)\,,\\
\begin{split}
P_{n+1}(x)=&-\int_{0}^{\ell-x}dy\,\varphi'(y)\,P_{n}(x+y)\\&-\int_{0}^{x}dy\,\varphi'(y)\,P_{n}(x-y)\,.
\end{split}
\end{align}
Nevertheless, we have found empirically that solving Eq.~\eqref{eq:pximplicit} self-consistently provides faster convergence and numerical stability~\footnote{Convergence is exponential with the number of iterations and independent of the initial guess for $P(x)$, which we take arbitrarily.} in comparison to the recursive solution~\eqref{eq:recursion}. We use the derived solution in Eq.~\eqref{eq:tcl}, to obtain the semi-classical expression of the transmittance. 

Using the newly found transmittance in formula ~\eqref{eq:LB}, we compute the associated semi-classical conductance $G_{\text{cl}}$.
In Fig.~\ref{fig:diffusive}, we compare $G_{\text{cl}}$ (solid
lines) with the exact quantum calculation $G_{\text{\ensuremath{\g}}}$
(dots) and find an excellent agreement for all chemical potentials and temperatures. 

Deep in the diffusive region, $\g N\gg1$, the semi-classical model
has some deviations with respect to the quantum solution. In the inset
of Fig.\ref{fig:diffusive}-bottom, we depict the relative error $\e_{r}=|G_{cl}-G_{\g}|/G_{\g}$ in the middle of the spectrum and verify it doesn't increase above 10\%.  One possible explanation for this discrepancy could be that the semi-classical model assumes that at each reset event the new momentum is drawn uniformly in the interval $[-\pi,\pi]$ and the particle has ballistic propagation at the corresponding velocity. In principle, we have to take into account the mode occupation of the fermions in the system. Indeed, the exclusion principle should prevent the particle to acquire a momentum corresponding to an already occupied mode. Taking these effects into account is however beyond the scope of this paper. 
We also stress that within this approach, we have considered leads and systems described by the same Hamiltonian in absence of dephasing. This assumption ensures that we do not need to take into account any additional reflection phenomena that might occur when the particle is transferred from the leads to the system.

The semi-classical picture provides an intuitive explanation of the conductance drop observed close to the band edges, $\mu_{\rm edge}=\pm 2t_x$. Close to these points, the velocity of incoming particles is the lowest. It is then more likely that a measurement process will occur and reset its speed, increasing its chance to backscatter into the original lead, and thus reducing the conductance. Additionally, the first measurement process resets the single-particle velocity, leading to a  uniform distribution of the particle over all the accessible states. 
Thus, after the measurement the particle attains an infinite temperature state, which is reservoir-independent and is the one related to the bulk transport properties described by the diffusion constant~\eqref{eq:D}. Remark that this picture is consistent with the fact that the diffusion constant evaluated in the bulk is independent of the boundary chemical potentials and temperatures. In conclusion, this simple semi-classical physical picture connects bulk and boundary effects on the transport properties of this system,  which are revealed by the diffusion constant and the conductance respectively.


\section{Dephased ladder \label{sec:Dephasingladder}}

We now extend the result for the conductivity of a 1D system to a ladder made of $M$ legs in the transverse direction, as described by the Hamiltonian~\eqref{eq:system}, see also Fig.~\ref{fig:model}. 
In this section, we consider noises that are site-to-site independent along the $x$ axis, but without
a fixed structure in the $y$ direction. Even though a natural choice is to consider noise processes which are  uncorrelated along the $y$-direction (as we will do), considering also correlated structures is motivated from synthetic dimensions setups. These setups make use of coupling between non-spatial degrees of freedom to simulate motion along  additional dimensions~\cite{Mancini1510,genkina2019imaging,chalopin_probing_2020,zhou2022observation}. In our setup, the $x$ direction would correspond to the physical dimension while the synthetic dimension is mapped to the transverse $y$ direction. With this mapping, the QSH studied here could be realized from randomly oscillating potentials that are spatially resolved in the physical direction, see also Sec.~\ref{sec:Conclusion}. 

We recall the generic expression for the QSH Eq.~\eqref{eq:deph_hamiltonian}
\begin{equation}\label{eq:noise2}
dH_{t}=\sqrt{2\gamma}\sum_{j,p,p'}a_{j,p}^{\dagger}a_{j,p'}dW_{t}^{j,p,p'}\,,
\end{equation}
with the covariance of the noise $dW_{t}^{j_{1},p_{1},p'_{1}}dW_{t}^{j_{2},p_{2},p'_{2}}=\delta_{j_{1},j_{2}}C_{p_{1},p'_{1},p_{2},p'_{2}}dt$. Each term of the sum describes a transition from a state indexed by $p'$ to a state indexed by $p$ with a random complex amplitude given by $dW_t^{j,p,p'}$. It is the most general way of writing noisy quadratic jump processes between different states in the $y$ direction. 

In what follows, we will investigate the transport for different geometries of the noise by specifying the covariance tensor $C$.

\begin{figure}
\begin{centering}
\includegraphics[width=1\columnwidth]{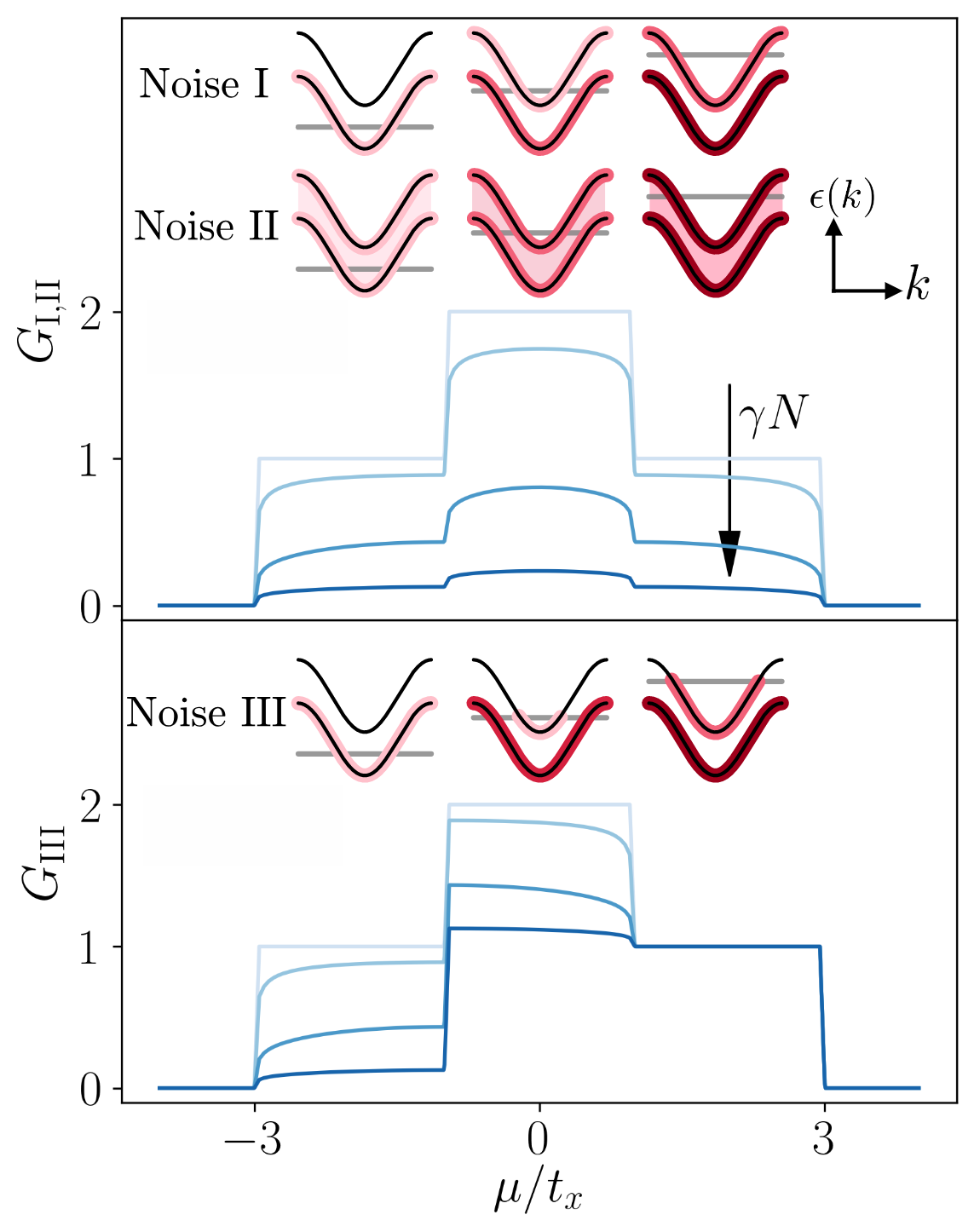}
\par\end{centering}
\caption{\label{fig:2-leg-ex}Conductance profiles for different correlations of the noise~\eqref{eq:noise2}
and increasing values of $\protect\g N$.  For increasing shades of blue: $\g=0$; $\g=0.1\,t_x,N=5$; $\g=0.5\,t_x,N=10$ and $\g=0.5\,t_x,N=50$ and $t_y=t_ x$. Noise I and II share the same conductance profile, while Noise III features the coexistence of ballistic and diffusive transport, see main text. The sketches on top of the conductance plots depict the stationary state reached in the bulk.  These states may or may not depend on the position of the chemical potential $\mu$ in the reservoirs (horizontal grey lines),  with respect to the dispersion relations of the different conduction modes in the system (black lines). The red halo on top of the dispersion relations indicates the occupation probability of the modes. Noise I distributes particle uniformly within each band separately, while Noise II distributes the states in all bands isotropically. Noise III is a special case which preserves the shape of the zero temperature distribution of the reservoirs in the bulk only in the upper band.}
\end{figure}


\subsection{Noise I}\label{sec:noiseI}
We start with  the simplest
case, that we label Noise I. It involves a single uniform noise
acting on a given vertical section of the system, see also Fig.~\ref{fig:model}-top. It is described by 
\begin{align}
\label{eq:noiseI}
\begin{split}
&dH_t^{\rm I}=\sqrt{2\gamma}\sum_{j,m}a_{j,m}^\dagger a_{j,m}dB_t^{j}=\sqrt{2\gamma}\sum_{j,p}a_{j,p}^\dagger a_{j,p} dB_t^{j}\,,
\end{split}\\
&C^{\rm I}_{p_1,p_2,p'_1,p'_2}=\delta_{p_1,p_2}\delta_{p_2,p'_1}\delta_{p'_1,p'_2}\,,
\end{align}  
with $\{B_{t}^{j}\}$ independent Brownian processes (recall that $m$ indexes the spatial degrees of freedom in the $y$ direction while $p$ indexes the transverse modes). 
This kind of noise can be naturally implemented in synthetic ladders generated from internal spin degrees of freedoms of ultracold atoms~\cite{Mancini1510,genkina2019imaging,chalopin_probing_2020,zhou2022observation}. Equation \eqref{eq:noiseI} would correspond to a randomly fluctuating potential that acts independently on each atom and uniformly shifts the energy levels of each spin state by $\sqrt{2\g}$.

The noise $dH_t^{\rm I}$ commutes at fixed $j$ with the occupation number operator of every mode $p$, $a_{j,p}^\dagger a_{j,p}$, and therefore does not couple different modes. As a consequence, all the results that we have derived for the conductance of a 1D system can be trivially extended to the present case since the system is then equivalent to a collection of uncoupled 1D bands. The dispersion associated to each band $\epsilon_p(k)$ is the same than for the 1D case with an overall energy shift given by  $\epsilon_p(\pi/2)$. Thus, the total conductance is the sum of the contribution of each mode, namely
\begin{equation}\label{eq:GI}
G_{\rm I}(\mu)=\sum_{p}G_{\gamma}\left[\mu-\epsilon_{p}\left(\frac\pi2\right)\right]\,,
\end{equation}
where $G_\gamma$ is given by Eq.~\eqref{eq:LB}, extensively studied in the purely 1D case.

In the absence of dephasing and at zero temperature $\gamma=T=0$), the conductance shows the usual staircase quantization  with respect to the chemical potential. As it is shown in Fig.~\ref{fig:2-leg-ex} for a 2-leg ladder, the jumps in conductance take place whenever the number of bands crossed by the chemical potential changes. For a finite rate $\g$, the action of the dephasing noise is the same for each individual band and, as a consequence, the total conductance decays as $1/N$ for larger systems.  

We stress that since Noise I doesn't mix the different modes, it cannot change the value of their occupation number, which is set by the chemical potential in the reservoirs. For instance, if a given mode was initially empty, it will remain so in the steady state. However, at fixed $p$, within \textit{a single band}, the dephasing noise~\eqref{eq:deph1D} drives the density matrix to a state proportional to the identity~\cite{CaiZi_AlgebraicvsExponentialDissipativesystems}, which is reminiscent of the infinite temperature state discussed in the 1D case, see also App.~\ref{App subsec: Heat}.  
We therefore say that Noise I is \textit{maximally mixing} the modes $k$ in the $x$ direction but not mixing at all the modes $p$ in the $y$ direction. As a consequence, it does not drive the system to a genuine infinite temperature state, this only happens within each individual band. A picture of this stationary state for increasing chemical potentials is sketched in Fig.~\ref{fig:2-leg-ex}.

We now show that the conductance profile illustrated in Fig.~\ref{fig:2-leg-ex} is not unique to the uncorrelated Noise I~\eqref{eq:noiseI}, and also describes other types of geometries.


\subsection{Noise II}

In this section, we consider an isotropic case, where the noise is uncorrelated both in the $x$ and $y$ directions. In this case, that we label Noise II, the QSH in position basis reads 
\begin{align}
dH_{t}^{{\rm II}}&=\sqrt{2\g} \sum_{j,m}c_{j,m}^{\dagger}c_{j,m}dB_{t}^{j,m}\,,
\end{align}
with $dB_{t}^{j_{1},m_{1}}dB_{t}^{j_{2},m_{2}}=\delta_{j_{1},j_{2}}\delta_{m_{1},m_{2}}dt$.  After performing the unitary transformation that diagonalizes the non-stochastic problem~\eqref{eq:system} in the form~\eqref{eq:systemdiag}, we find the noise correlation function
\begin{align}
C_{p_{1},p'_{1},\atop p_{2},p'_{2}}^{{\rm II}}&=\left(\frac{2}{M+1}\right)^{2}\sum_{m=1}^{M}\prod_{a=p_{1},\atop p'_{1},p_{2},p'_{2}}\sin\left(\frac{\pi a m }{M+1}\right)\,.
\end{align}
Contrary to the previous case, Noise II is maximally mixing for the modes $k$ in the $x$ direction and for the modes $p$ in the $y$ direction. As a consequence, this noise drives the system to a genuine infinite temperature state in the bulk, see also sketches in Fig.~\ref{fig:2-leg-ex}. The mixing of the modes in the transverse direction renders the task of computing the conductance an a priori non-trivial one.

Nevertheless, as we show in App.~\ref{app:proofcond}, for any noise satisfying the condition
\begin{eqnarray}\label{eq:condition}
\sum_{p}C_{p_{1},p,p,p_{2}} & = & {\cal N}\delta_{p_{1},p_{2}}\,,
\end{eqnarray}
the equations of motion of the total current $J$ coincide to those generated by Noise I up to a renormalization of $\gamma$ by a constant ${\cal N}$. Using that $\delta_{a,a'}=\frac{2}{M+1}\sum_{j}\sin(\frac{\pi aj}{M+1})\sin(\frac{\pi a'j}{M+1})$, one can verify that $C^{\rm II}$ satisfies condition~\eqref{eq:condition} with  $\mathcal{N}=1$.
Thus we find that 
\begin{equation}\label{eq:GIGII}
G_{{\rm I}}(\mu)=G_{{\rm II}}(\mu)\,.
\end{equation}
This result may sound surprising as, even though Noise
II drives the system towards the maximally mixed, infinite temperature state, the
staircase behavior of $G$ is preserved, {\it i.e.} there is a discontinuity of $\partial_{\mu}G$ every time the chemical potential touches a band. 

The remarkable equality~\eqref{eq:GIGII} can be intuitively understood within the semi-classical picture. All that matters for the conductance is the number of modes that can contribute to the current. This number is fixed by the chemical potential, which in turn controls the staircase behavior of the conductance. Once a particle has entered the system, different scattering events may switch it from one channel to the other isotropically, as expressed mathematically by the condition~\eqref{eq:condition}. Nevertheless, all the channels carry the current in the same fashion, since the dispersion relations of all the transverse modes coincide in quasi-momentum $k$, except for an irrelevant energy shift. As a consequence, the total conductance is insensitive to whether the noise is coherent (or not) along the transverse direction. 

\subsection{Noise III}

Finally, we illustrate how breaking the condition~\eqref{eq:condition} may lead to exotic transport.
We introduce the case of Noise III, where the correlations $C^{\rm III}$ of Noise III are designed such that they only couple pairs of transverse modes:
\begin{equation}
C_{p_{1},p'_{1},p_{2},p'_{2}}^{{\rm III}}=f_{p_{1},p_{2}}\delta_{p_{1},p'_{2}}\delta_{p'_{1},p_{2}} \,,
\end{equation}
for which 
\begin{align}
\sum_{p}C_{p_{1},p,p,p_{2}}^{{\rm III}} & =\delta_{p_{1}p{}_{2}}\sum_{p}f_{p_{1},p}\,.
\end{align}
The isotropy condition \eqref{eq:condition} is fulfilled if, per example, we impose $f_{p_1,p_2}$ to be equal to a constant $c$ for every $(p_1,p_2)$ in which case we have that ${\cal N}=Mc$.

Breaking the isotropy condition can lead to a hybrid type of transport. For instance, let us consider the case where 
\begin{equation}
f_{p_{1},p_{2}}=\delta_{p_{1},p_{2}}\theta(p-p_0)\,,
\end{equation}
with the convention $\theta (0)=0$ for the Heaviside step function.
This noise
imposes diffusive transport to the lowest transverse modes
($p\leq p_{0}$) while the highest modes ($p>p_{0}$) remain
ballistic. Since this noise does not couple the different transverse
modes, the conductance has both ballistic and diffusive contributions:
\begin{equation}
G_{\text{III}}(\mu)=\sum_{p\leq p_{0}}G_{\g}(\mu-\epsilon_{y,p})+\sum_{p>p_{0}}G_{\g=0}(\mu-\epsilon_{y,p}).
\end{equation}
We plot an example of such situation on Fig.\ref{fig:2-leg-ex}-bottom
for a 2-leg ladder and $p_{0}=1$. The overall current has diffusive
behavior until the chemical potential reaches the bottom of the upper band at $\mu_{0}=-2t_{x}-2t_{y}\cos\left(\frac{p_{0}\pi}{M+1}\right)$.
For $\mu\geq\mu_0$ the ballistic mode starts contributing and dominates the conductance in the thermodynamic limit $N \ra \infty$


\section{Dephased leads\label{sec:Out-of-equilibrium-leads}}

Lastly, we discuss how to describe the case where the leads are themselves affected by the dephasing noise. The goal of this section is to describe transport deep inside the system where subportions of the system act as effective reservoirs.

\subsection{1D case}
We again start by discussing in detail the simpler 1D case.
Recall that from Eq.~\eqref{eq:self-leads}, the reservoirs contribution to the self-energy was given by the retarded components of their Green's function. For the left bath, we had $\Sigma_{{\rm L},i,j}^{R/A}=t_{x}^{2}g_{0,0}^{R/A}\delta_{i,j}\delta_{i,1}.$ From Eq.~\eqref{eq:self-energy} one sees that adding a local dephasing term $dH_{t}=\sum_{j}c_{j}^{\dagger}c_{j}dB_{t}^{j}$ on all the sites of the leads gives a contribution to the leads self-energy given by $\Sigma_{j,j'}^{R/A}(t,t')=\pm i\frac{\gamma}{2}\delta_{j,j'}\delta(t-t')$ or $\Sigma_{k,k'}^{R/A}(\omega)=\pm i\frac{\gamma}{2}\delta_{k,k'}$ in momentum and frequency space. From the Dyson equation for the retarded and advanced part, $G_{k,k'}^{R/A}=([g_{k,k'}^{R/A}]^{-1}-\Sigma_{\gamma,k,k'}^{R/A})^{-1}=\delta_{k,k'}(\omega-\epsilon_{k}\pm i\frac{\gamma}{2})^{-1}$. This means that one can obtain the retarded and advanced Green's functions of the leads affected by dephasing by simply shifting the frequency $\omega\to\omega\pm i\frac{\gamma}{2}$. The expressions of the Green's function $g_{0,0}^{R/A}$ and $g_{N+1,N+1}^{R/A}$ in the absence of dephasing are given in App.~\ref{App subsec: Green'sfunction} and their shift in frequency leads to the following contribution of the leads to the self-energy of the system : 
\begin{equation}\label{eq:self-lead-dif}
\Sigma_{{\rm L/R,}i,j}^{R/A}	=\delta_{i,j}\delta_{j,1/N}\frac{1}{2}\left(\omega \pm i\frac{\gamma}{2}\mp i\sqrt{4t_{x}^{2}-\left(\omega \pm i\frac{\gamma}{2}\right)^{2}}\right).
\end{equation}
The Keldysh component can be taken to be in local equilibrium in the leads as shown in App.~\ref{App subsec: Heat}. Indeed, as stated before, the dephasing noise \eqref{eq:deph1D} drives the system towards a maximally mixed state.
It can be parametrized by a Gibbs state with an infinite temperature and chemical potential such that the ratio $\mu/T$ is fixed to match a local occupation number imposed by the lead. The stationary state reached by the lead is thus uniquely
characterized by the particle density $n_{{\rm L/R}}$ and fulfills
the fluctuation dissipation relation $g^{K}(\omega)=2i(1-2n_{\text{L/R}})\Im\left(g^{R}\right)$,
which in turns gives for the contribution of the leads  to the self-energy of the system  
\begin{equation}
\S_{\text{L/R}}^{K}(\omega)=2i(1-2n_{{\rm L/R}})\Im\left(\S_{\text{L/R}}^{R}(\w)\right).\label{eq:effectiveGreenK}
\end{equation}
Under this protocol, the imbalance
between the density of the leads $\Delta n=n_{{\rm L}}-n_{{\rm R}}$ is responsible for driving the current. In the large system size limit, the current is directly proportional to the density imbalance as expected from Fick's law 
\begin{equation}
J=-\frac{D\Delta n}{N}.
\end{equation}
This result originates from the numerical solution of Eq.~\eqref{eq:self-energy} which confirms that the density profile is linear with the proportionality coefficient (diffusion constant) given by Eq.~\eqref{eq:D}, see also previous works \citep{Znidaric__XXdeph,BauerBernardJin_Stoqdissipative,Jin_Quantumresistors}. 

\subsection{$M$-leg ladder}

We briefly discuss the case of the $M$-leg ladder in the case of Noise
I from Eq.~\eqref{eq:noiseI}. As in the one-dimensional case discussed above,  Eq.~\eqref{eq:effectiveGreenK} remains valid for each individual band, since the noise does not mix the different transverse momenta sectors.

As before, the total current results from 
the sum of the current of each transverse mode. In turn, these are determined
by the particle number $n_{{\rm L/R}}$ of the leads, which
depends on the initial state chosen
\begin{align}
J & =\sum_{p}J_{p} =-\frac{D}{N}\Delta n_{{\rm tot}}\,.
\end{align}
Importantly, we see that all the channels contribute in the same way
meaning that the total current is just equal to the total density
imbalance, regardless of which mode is filled. In stark contrast with the previous section, the current remains the same for all values
of the filling of the boundary leads, only the relative imbalance in the total occupation numbers
matters.

\subsection{Markovian limit}

In general, coupling a lead to a system induces non-trivial
memory effects as indicated by the frequency dependence of the self
energy $\S_{\text{L/R}}$ in Eq.\eqref{eq:self-lead-dif}. A possible
limit to create a Markovian lead is to take large dephasing
rates $\gamma\gg t_{x}$, leading to  
\begin{align}
&\S^{R/A}_{{\rm L/R},i,j}(\omega) =\mp i\delta_{i,j}\delta _{j,1/N}\frac{2t_{x}^2}{\gamma}, \nonumber\\
&\S^{K}_{{\rm L/R},i,j}(\omega) =-i\frac{4}{\gamma}t_{x}^{2}(1-2n_{{\rm \text{L/R}}})\delta_{i,j}\delta_{1/N}\,,
\end{align}
in which case the frequency dependence vanishes. In Ref.~\cite{JinFilipponeGiamarchi_GenericMarkovian}, it was shown
that such Markovian bath is equivalent to coupling the system to a
Linblad operator
\begin{align}
{\cal L}(\rho)= & 2\alpha\left(c^{\dagger}\rho c-\frac{1}{2}\{(1-c^{\dagger}c),\rho\}\right)\nonumber \\
 & +2\beta\left(c\rho c^{\dagger}-\frac{1}{2}\{c^{\dagger}c,\rho\}\right)
\end{align}
with $c^{\dag}$ the creation operator at the coupling site, $\alpha$ the injecting rate  and $\beta$ the extracting rate 
\begin{equation}
\alpha=n_{\text{L/R}},\quad\beta=1-n_{\text{L/R}}
\end{equation}
whose action is to fix a target density $n_{\text{L/R}}$ on the site it is coupling to. As before, $n_{\text{L/R}}$ is completely determined
by the initial state of the lead before coupling.

For an $M$-leg ladder, we will end up with $M$ Lindbladians, one for each mode, and whose injecting rate will be fixed by the occupation number of the mode.


\section{Conclusion and perspectives\label{sec:Conclusion}}
In this work, we have studied the current flowing through $M$-leg
ladders subject to external dephasing noises with different correlations along the direction $y$ transverse to transport. Starting from the purely one-dimensional case, $M=1$, we have devised a semi-classical model to compute the conductance as a function of the chemical potential and found excellent agreement with numerical solutions obtained from exact self-consistent calculations of Dyson's equation. 

Showing the effectiveness of this semi-classical model is important as it allows to build a simple and intuitive physical picture of the emergence of bulk resistive behavior in quantum stochastic resistors. As extensively discussed in the core of this paper, the bulk transport properties of these systems, such as the diffusion constant or the resistivity, are insensitive to the temperature and chemical potential of the connected reservoirs. We showed that this is not the case for their conductance and that we could rely on the semi-classical approach to bridge between boundary and bulk effects. It could be interesting to understand the deeper connections between our semi-classical model and the quasi-particle picture recently introduced to describe entanglement growth~\citep{Cao_2019,Turkeshi2021}.

We have also shown that these non-trivial results in one-dimension could be also extended to $M$-leg ladder systems. In particular, we have shown that the results valid in 1D could be immediately applied to the case where the noise term preserves the coherence in the vertical direction, this case being particularly relevant to systems featuring synthetic dimensions~\cite{Mancini1510,genkina2019imaging,chalopin_probing_2020,zhou2022observation}. In this case, the total conductance is just the sum of the contributions of independent 1D channels and its diffusion constant remains unchanged.

We then demonstrated that the coherence properties of the noise along the $y$ direction do not play a role on the conductance of a ladder system when the noise fulfills the condition~\eqref{eq:condition}. We also showed that breaking this condition for the correlations of the noise allows to engineer exotic transport. We gave an example (Noise III) where the longitudinal current switches between a diffusive or ballistic behavior depending on the chemical potential. 

Lastly, we studied the case where the noise
acts on the leads themselves. For leads affected by the coherent
Noise I, we showed that the current was only dependent on the density difference 
between left and right leads, independently of the absolute value of these fillings. We also showed that in the large dephasing limit $\gamma \gg t_x$, the reservoirs could be effectively described as Markovian injectors. 

In the latter case, the contribution of each mode of the current was
the same. This raises the natural question of understanding what would
happen if this degeneracy were to be lifted. A particularly interesting
problem would be to understand the effect of density-density interactions
in the transverse direction to the transport. In the two-leg ladder, numerical
studies relying on DMRG techniques could be supported by the infinite
system size perturbation technique recently introduced in \citep{Jin_Quantumresistors}. 

We briefly comment on the prospect of experimental realizations.
The noise discussed here is specially suitable for implementation
in synthetic dimensions setups such as ultracold atoms in shaken
constricted optical channels~\citep{salerno_quantized_2019} or with synthetic spin dimension~\citep{Mancini1510,genkina2019imaging,chalopin_probing_2020,zhou2022observation,livi2016synthetic}, or even photonic systems with ring resonator arrays~\citep{ozawa2016synthetic,mittal2019photonic}. In these
systems, the synthetic dimension plays the role of transverse direction
in our model while the physical dimension encodes the longitudinal
direction. A dephasing noise in the physical dimension would thus
affect in the same manner all the synthetic sites, giving a natural
realization of Noise I described by Eq.~\eqref{eq:noiseI}.


\begin{acknowledgments} This work has been supported
by the Swiss National Science Foundation under Division II. J.S.F.
and M.F. acknowledge support from the FNS/SNF Ambizione Grant No.
PZ00P2\_174038.
\end{acknowledgments}

\appendix

\section{Green's function of the free system \label{App subsec: Green'sfunction}}

To compute the current operator~\eqref{eq:current}, one first needs to compute
the Green's function of the lead alone and of the system in presence
of the leads. For simplicity, we treat here the 1D channel but generalization
to $M$-legs will be straightforward. For a single left lead, the Hamiltonian ~\eqref{eq:system} can be divided as 
\begin{equation}
H=H_{{\rm S}}+H_{{\rm L}}-t_{x}(c_{0}^{\dagger}c_{1}+c_{1}^{\dagger}c_{0}).
\end{equation}
We suppose the lead (L) and the system (S) to be non-interacting so that $H_{{\rm L,S}}$ is a quadratic
Hamiltonian. The associated action in the Keldysh formalism, using
Larkin notation \citep{Larkin_vortices_supra} for the fermionic fields, 
\begin{align}
S= & S_{{\rm S}}+\int\frac{d\w}{2\pi}\left([\bar{\psi}_{\rm L}]\boldsymbol{g}^{-1}_{\rm L}[\psi_{\rm L}]\right)\nonumber \\
 & +t_{x}\left(\bar{\psi}_{{0}}^{1}\psi_{1}^{1}+\bar{\psi}_{{0}}^{2}\psi_{1}^{2}+\bar{\psi}_{{1}}^{1}\psi_{0}^{1}+\bar{\psi}_{{1}}^{2}\psi_{0}^{2}\right)
\end{align}
where $S_{{\rm S}}$ is the action of the system, $[\psi_{{\rm L}}]$
is a vector containing all Grassman variables associated to the left lead
and $\boldsymbol g$ is  the Green's function before coupling, with the same matrix structure as $\boldsymbol{\mathcal{G}}$ in Eq.~\eqref{eq:Dyson},  $\boldsymbol{g}:=\begin{pmatrix}g^{R} & g^{K}\\
0 & g^{{\rm A}}
\end{pmatrix}$. Integrating out the lead's degrees
of freedom, one finds 
\begin{align}
&S=S_{{\rm S}}-\int\frac{d\w}{2\pi}\begin{pmatrix}\bar{\psi}_{{1}}^{1} & \bar{\psi}_{{1}}^{2}\end{pmatrix}\begin{pmatrix}\Sigma^{R} & \Sigma^{K}\\
0 & \Sigma^{A}
\end{pmatrix}\begin{pmatrix}\psi_{1}^{1}\\
\psi_{1}^{2}
\end{pmatrix}, \nonumber\\
&\Sigma^{R/A/K} =t_{x}^{2}g_{0,0}^{R/A/K}\,.
\end{align}

For an infinite-size lead made of  $n$ discrete sites with a tight-binding
Hamiltonian coinciding with Eq.~\eqref{eq:system}, the spectrum is given by $\epsilon_{k}=-2t_{x}\cos\left(\frac{k\pi}{n+1}\right),\quad k\in[1,n]$.
The associated retarded Green's function in momentum space is given by (the tilde designates momentum space) 
\begin{equation}
\tilde{g}_{k,k'}^{R}(\omega)=\frac{\delta_{k,k'}}{\omega+2t_{x}\cos\left(\frac{k\pi}{n+1}\right)+i0^{+}}\,.
\end{equation}
In position space, this yields 
\begin{align}
g_{j,j'}^{R}= & \frac{2}{n+1}\sum_{k}\sin\left(\frac{k(j+1)\pi}{n+1}\right)\sin\left(\frac{k(j'+1)\pi}{n+1}\right)\nonumber \\
 & \frac{1}{\omega+2t_{x}\cos\left(\frac{k\pi}{n+1}\right)+i0^{+}}\,.
\end{align}
We are interested in the $j=j'=0$ term in the semi-infinite limit,
i.e. we take $n\to\infty$. Introducing $p=\frac{k\pi}{n+1}$, we get
\begin{equation}\label{eq:g00r}
g_{0,0}^{R}=\frac{2}{\pi}\int_{0}^{\pi}dp\frac{\sin^{2}p}{\left(\omega+2t_{x}\cos p+i0^{+}\right)}\,.
\end{equation}
which can be computed by contour integral in the complex plane to
be:
\begin{equation}\label{eq:grl11}
g_{0,0}^{R}=\frac{1}{2t_{x}^{2}}\left(\omega+i0^+-i\sqrt{(2t_{x})^{2}-(\omega+i0^+)^{2}}\right) \,.
\end{equation}
The advanced component is just the complex conjugate of the retarded one. To obtain the Keldysh component, we will suppose that the lead
is at thermal equilibrium so that 
\begin{align}
 & g_{0,0}^{K}(\omega)\nonumber \\
 & =\tanh\left(\frac{\omega-\mu}{2T}\right)2i\im\left(g_{0,0}^{R}\right),\nonumber \\
 & =
-\theta(2t_x-|\omega|)\frac{i}{t_{x}^{2}}\tanh\left(\frac{\omega-\mu}{2T}\right)\sqrt{(2t_{x})^{2}-\omega^{2}}
\end{align}
which is enough to compute the Green's function of the system in presence of the leads \cite{usmani1994inversion}. To obtain the result with dephasing in the leads, one should replace $0^+\rightarrow \g/2$.

In the absence of noise, the Green's function of the system is easily computed by noticing
that the system with both leads constitutes a discrete tight-binding
chain of infinite size. In this case, the Green's function in momentum space is given by 
\begin{equation}
\tilde{\mathcal{G}}^{R}(p,p')=\frac{\delta(p-p')}{\omega+2t_{x}\cos p+i0^{+}}.
\end{equation}
By doing the inverse Fourier transform we get it in position space
\begin{align}
\mathcal{G}_{j,k}^{R} & =\int_{-\pi}^{\pi}\frac{dp}{2\pi}\frac{e^{-ip(j-k)}}{\omega+2t_{x}\cos p+i0^+}
\end{align}
which can be again computed by contour integral to be  : 
\begin{equation}
\mathcal{G}_{j,k}^{R}=\frac{z_1^{k-j}}{z_1-z_2}
\end{equation}
for $j\leq k$ with $z_{1,2}:=-\left(\frac{\omega+i0^{+}}{2t_{x}}\right)\mp i\sqrt{1-\left(\frac{\omega+i0^{+}}{2t_{x}}\right)^{2}}$.
Using the symmetry property $\mathcal{G}_{j,k}^{R}=\mathcal{G}_{k,j}^{R}$ we have the
full Green's function in position space.

\section{ Re-scaled conductance profiles
\label{App subsec: Conductance}
}

\begin{figure}
\begin{centering}
\includegraphics[width=1\columnwidth]{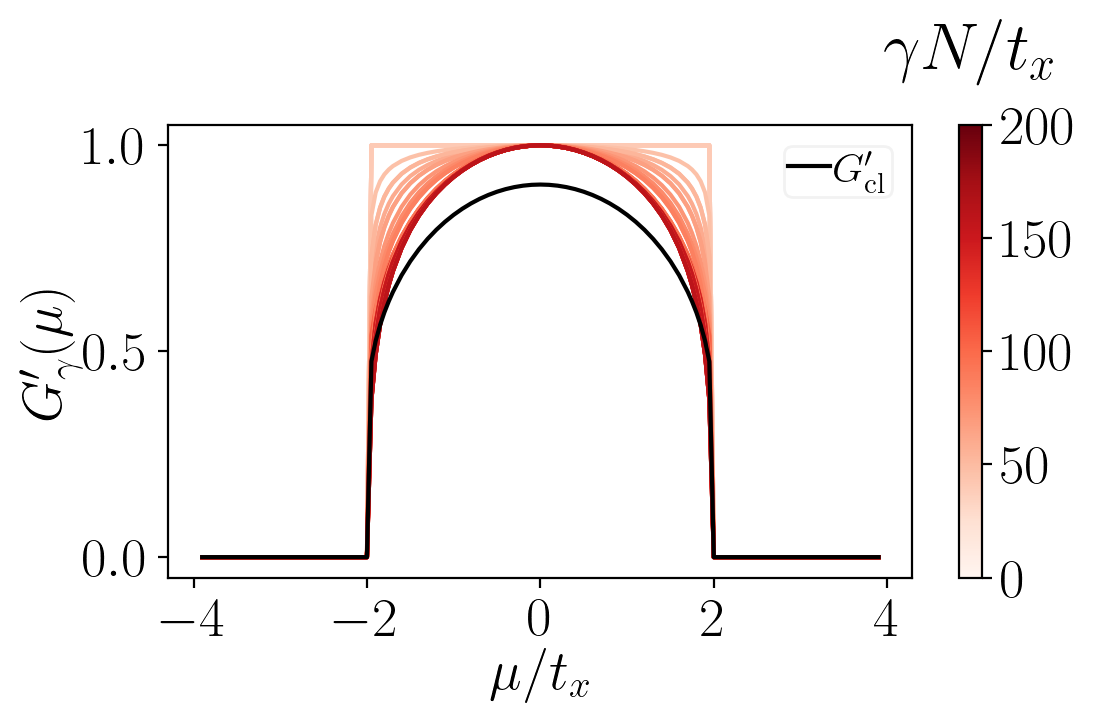}
\par\end{centering}
\caption{\label{fig:dome}Rescaled conductance profiles $G'_{\g}(\mu,N):=G_{\g}(\mu,N)/G_{\g}(0,N)$ for different values of $\g N$ with $\g=[0,2]t_x$ and $N=[5,200]$. The rescaled conductance converges in the diffusive limit $\g N \gg 1$ to a curve qualitatively similar to the semi-classical expectations in the same limit.}
\end{figure}

In this section, we present further numerical data on the conductance profiles in the diffusive regime. In this regime, the conductance decays with the inverse system size $G \propto 1/N$ thus, to focus on the chemical potential dependence, we depict in Fig.~\ref{fig:dome} the rescaled conductance $G'_{\g}(\mu,N):=G_{\g}(\mu,N)/G_{\g}(0,N)$ profiles for different values of $\gamma N$  and compare with the semi-classical rescaled value $G'_{\g,cl}(\mu,N)=G_{\g,cl}(\mu,N)/G_{\g}(0,N)$. The dependence of $G_\g(\mu=0)$ on the system size $N$ is plotted in Fig.~\ref{fig:diffusive}. As we transition to the diffusive limit, the rescaled curves converge to a dome-like shape which follows the qualitative dependence of the semi-classical approach, see black line for $G'_{\mathrm{cl}}(N \ra \infty)$. The deviations between the exact and semi-classical approach are more significant in the center of the band but never exceed $10\%$.

We note that such strong dependence on the thermodynamic properties of the leads is not present in the diffusion constant and is a unique property of the conductance.

\section{ Mixing effects with dephasing noise
\label{App subsec: Heat}
}

\begin{figure}
\begin{centering}
\includegraphics[width=1\columnwidth]{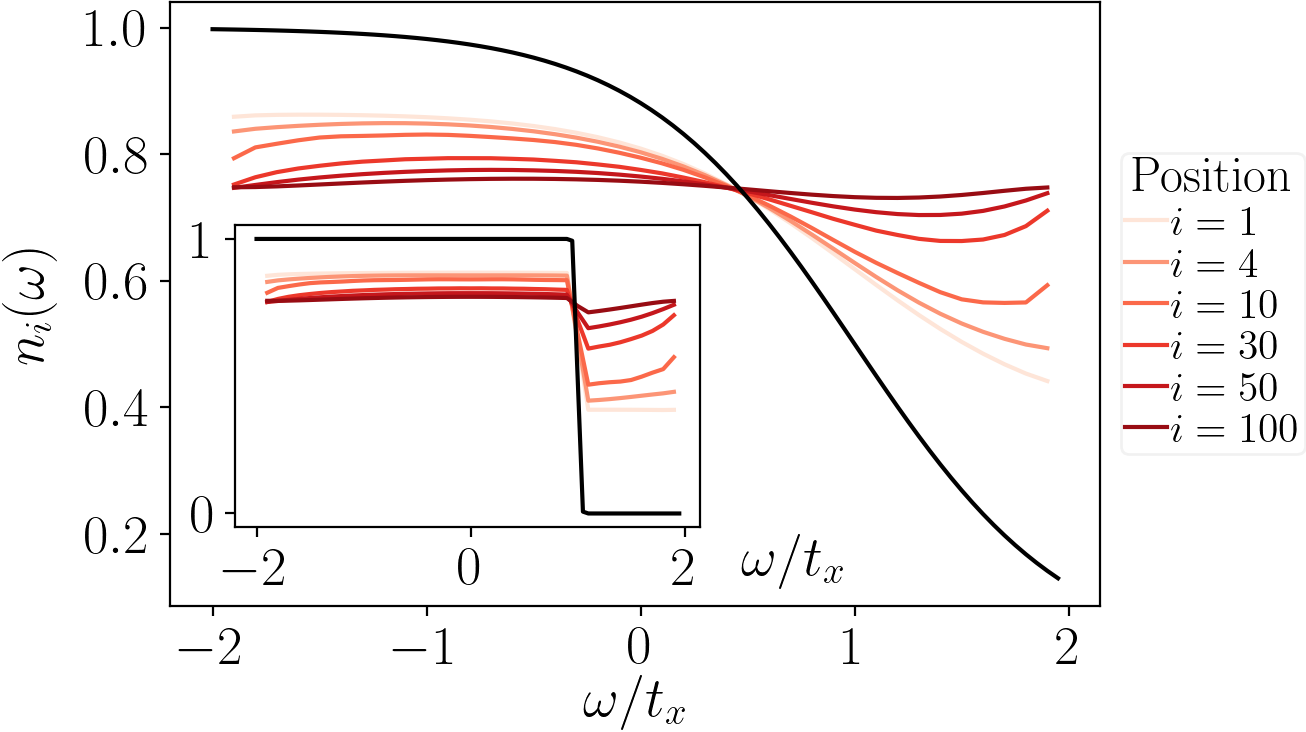}
\par\end{centering}
\caption{\label{fig:heat}Non-equilibrium local occupation distribution extracted from Eq.~\eqref{eq:n(w)}  at different positions in a chain of $N=200$ sites with on-site dephasing rate $\g=0.05t_x$.  The chain is coupled on the first and last site $i=1,200$ to a thermal lead with $\mu=t_x$ and $T=0.5t_x$. The black line corresponds to the Fermi distribution of the attached reservoirs. Inset: same plot but the temperature of the leads is $T=0.01t_x$.}
\end{figure}

In this appendix, we discuss the stationary state induced by dephasing noise on a one-dimensional tight-binding chain. In the absence of dephasing, the whole chain will be in thermal equilibrium with a chemical potential and temperature matching the lead. Locally, it implies that the Green functions satisfy the fluctuation dissipation relation:

\begin{equation}
    \mathcal{G}^K_{i,i}(\omega)=(1-2n_i(\omega))\left(\mathcal{G}^R_{i,i}(\w)-\mathcal{G}^A_{i,i}(\w)\right) \label{eq:n(w)}
\end{equation}
with $n_i(\w)$ the local Fermi distribution with parameters $\mu,T$.

Beyond thermal equilibrium, we can still use Eq.~\eqref{eq:n(w)} as an ad-hoc definition of $n_i(\w)$ to characterize local deviations from equilibrium. 

The presence of any dephasing rate drives the system out of equilibrium. As explained in the main text, the dephasing maximally mixes the longitudinal momentum states. Since these, in 1D, label all the eigenstates of the system th action of dephasing corresponds to heat the system to infinite temperature. However, since the dephasing terms commute with the local particle number operator, the attained steady-state preserves a well defined particle number. In other words, the local density matrix deep into a dephasing region resembles a thermal distribution with an effective $\mu^*$ and $T^*$ such that $\mu^*,T^*\rightarrow \infty$ and $\mu^*/T^*$ tuned in such a way to have in the system the same spatial averaged particle density than the one in the attached leads. 

This is clear in Fig.~\ref{fig:heat} where we plot $n_i(\w)$ for different points in a dephasing chain coupled to a thermal lead on the left and right. By definition, the lead is in local thermal equilibrium and $n_{\rm L}(\w)$ is given by the Fermi-distribution, see black line. Near the lead, $n(\w)$ deviates strongly from a thermal distribution indicating that the system is far from equilibrium.
Deep into the chain, that is at distances from the leads larger than the scattering length ($i>t_x/\g$, in Fig.~\ref{fig:heat}), $n(\w)$ becomes flat as expected from a state with infinite $\mu,T$. The exact ratio $\mu^*/T^*$ is uniquely determined from the density of the reservoirs but its exact value depends on the distribution of the system near the edges.


\section{Proof of the condition \eqref{eq:condition}}\label{app:proofcond}

In this appendix, we give a proof of the condition Eq.~\eqref{eq:condition}
in the main text. The strategy is to write down the equations of motion
for the total current and derive a condition under which they are equivalent (up to a factor) for different types of noise.

The action of the deterministic part $H$ for the total current \eqref{eq:current} evaluated at site $j$, is given for any site by 
\begin{equation}
     \begin{split}
 \partial_tJ=i[H,J]=&  it_{x}^{2}\sum_{p}\Big[(n_{j+1,p}-n_{j,p})  \\
  &+a_{j-1,p}^{\dagger}a_{j+1,p}-a_{j,p}^{\dagger}a_{j+2,p} \\ &-a_{j+2,p}^{\dagger}a_{j,p}+a_{j+1,p}^{\dagger}a_{j-1,p}\Big]\,. 
  \end{split}
\end{equation}
Since $H$ is quadratic, and since it doesn't mix the different
modes by construction, its further action on quadratic operators will
only generate terms of the type $a_{j+k,p}^{\dagger}a_{j+k,p'}$.
By construction,  the QSH \eqref{eq:deph_hamiltonian} conserves the total number of particles at a given position
on the $x$ axis, i.e $[dH_t,n_{j,{\rm tot}}]=0$. Then,
one sufficient but not necessary condition for the equations of motion to have
the same form for all protocols is that the dual action on operators $\mathcal{L}^*$ for the averaged noise does not produce any new terms, i.e we must have for $j\neq j'$
\begin{equation}
{\cal L}^{*}(a_{j,p}^{\dagger}a_{j',p})=-{\cal N}a_{j,p}^{\dagger}a_{j',p}\,,\label{eq:sufficientcondition}
\end{equation}
where ${\cal N}$ is a constant which depends on the type of noise we are interested in. Invoking the
locality of the noise operator with respect to the longitudinal direction
we have that 
\begin{equation}
{\cal L}^{*}(a_{j,p}^{\dagger}a_{j',p})={\cal L}^{*}(a_{j,p}^{\dagger})a_{j',p}+a_{j,p}^{\dagger}{\cal L}^{*}(a_{j',p})\,.
\end{equation}
So the sufficient condition (\ref{eq:sufficientcondition}) can be cast into an even more restrictive one where we impose that $\forall p$, $\{a_{j,p}^{\dagger},a_{j,p}$\}
are eigenvectors of the operator ${\cal L}^{*}$. 

Recall the explicit expression of $\mathcal{L}^*$ 
\begin{align}
{\cal L}^{*}(\hat{O})= & \gamma\sum_{j,p_{1,2},p'_{1,2}}C_{p_{1},p'_{1},p_{2},p'_{2}}(2a_{j,p_{1}}^{\dagger}a_{j,p'_{1}}\hat{O}a_{j,p_{2}}^{\dagger}a_{j,p'_{2}}\nonumber \\
 & -\{a_{j,p_{1}}^{\dagger}a_{j,p'_{1}}a_{j,p_{2}}^{\dagger}a_{j,p'_{2}},\hat{O}\})\,.
\end{align}
 Using that $C_{p_{1},p'_{1},p_{2},p'_{2}}=C_{p'_{1},p_{1},p'_{2},p_{2}}$, we have that 
\begin{align}
{\cal L}^{*}(a_{j,p}^{\dagger}) & =-\gamma \sum_{p_{1},p'}C_{p_{1},p',p',p}a_{j,p_{1}}^{\dagger},\\
{\cal L}^{*}(a_{j,p}) & =-\gamma \sum_{p_{1},p'}C_{p',p_{1},p,p'}a_{j,p_{1}}\,,
\end{align}
and a sufficient condition for $\{a_{j,p}^{\dagger},a_{j,p}$\} to be
eigenvectors of ${\cal L}^{*}$ is then 
\begin{eqnarray}
\sum_{p}C_{p_{1},p,p,p_{2}} & = & {\cal N}\delta_{p_{1},p_{2}}\,,\label{eq:condition-1}
\end{eqnarray}
which is Eq.~\eqref{eq:condition} of the main text.

For the coherent Noise I, one has ${\cal N}=1$ and the transport properties
of a given protocol model can be deduced from those of Noise I by rescaling the coefficient $\gamma\to{\cal N}\gamma$.

\bibliographystyle{apsrev4-2}

\bibliography{biblio_all}

\begin{thebibliography}{72}%
\makeatletter
\providecommand \@ifxundefined [1]{%
 \@ifx{#1\undefined}
}%
\providecommand \@ifnum [1]{%
 \ifnum #1\expandafter \@firstoftwo
 \else \expandafter \@secondoftwo
 \fi
}%
\providecommand \@ifx [1]{%
 \ifx #1\expandafter \@firstoftwo
 \else \expandafter \@secondoftwo
 \fi
}%
\providecommand \natexlab [1]{#1}%
\providecommand \enquote  [1]{``#1''}%
\providecommand \bibnamefont  [1]{#1}%
\providecommand \bibfnamefont [1]{#1}%
\providecommand \citenamefont [1]{#1}%
\providecommand \href@noop [0]{\@secondoftwo}%
\providecommand \href [0]{\begingroup \@sanitize@url \@href}%
\providecommand \@href[1]{\@@startlink{#1}\@@href}%
\providecommand \@@href[1]{\endgroup#1\@@endlink}%
\providecommand \@sanitize@url [0]{\catcode `\\12\catcode `\$12\catcode
  `\&12\catcode `\#12\catcode `\^12\catcode `\_12\catcode `\%12\relax}%
\providecommand \@@startlink[1]{}%
\providecommand \@@endlink[0]{}%
\providecommand \url  [0]{\begingroup\@sanitize@url \@url }%
\providecommand \@url [1]{\endgroup\@href {#1}{\urlprefix }}%
\providecommand \urlprefix  [0]{URL }%
\providecommand \Eprint [0]{\href }%
\providecommand \doibase [0]{https://doi.org/}%
\providecommand \selectlanguage [0]{\@gobble}%
\providecommand \bibinfo  [0]{\@secondoftwo}%
\providecommand \bibfield  [0]{\@secondoftwo}%
\providecommand \translation [1]{[#1]}%
\providecommand \BibitemOpen [0]{}%
\providecommand \bibitemStop [0]{}%
\providecommand \bibitemNoStop [0]{.\EOS\space}%
\providecommand \EOS [0]{\spacefactor3000\relax}%
\providecommand \BibitemShut  [1]{\csname bibitem#1\endcsname}%
\let\auto@bib@innerbib\@empty
\bibitem [{\citenamefont {Akkermans}\ and\ \citenamefont
  {Montambaux}(2007)}]{akkermans_mesoscopic_2007}%
  \BibitemOpen
  \bibfield  {author} {\bibinfo {author} {\bibfnamefont {E.}~\bibnamefont
  {Akkermans}}\ and\ \bibinfo {author} {\bibfnamefont {G.}~\bibnamefont
  {Montambaux}},\ }\href {https://doi.org/10.1017/cbo9780511618833} {\emph
  {\bibinfo {title} {Mesoscopic Physics of Electrons and Photons}}}\ (\bibinfo
  {publisher} {Cambridge University Press},\ \bibinfo {year}
  {2007})\BibitemShut {NoStop}%
\bibitem [{\citenamefont {Giamarchi}(1991)}]{giamarchi_umklapp_1d}%
  \BibitemOpen
  \bibfield  {author} {\bibinfo {author} {\bibfnamefont {T.}~\bibnamefont
  {Giamarchi}},\ }\href {https://doi.org/10.1103/PhysRevB.44.2905} {\bibfield
  {journal} {\bibinfo  {journal} {Phys. Rev. B}\ }\textbf {\bibinfo {volume}
  {44}},\ \bibinfo {pages} {2905} (\bibinfo {year} {1991})}\BibitemShut
  {NoStop}%
\bibitem [{\citenamefont {Rosch}\ and\ \citenamefont
  {Andrei}(2000)}]{rosch_conservation_1d}%
  \BibitemOpen
  \bibfield  {author} {\bibinfo {author} {\bibfnamefont {A.}~\bibnamefont
  {Rosch}}\ and\ \bibinfo {author} {\bibfnamefont {N.}~\bibnamefont {Andrei}},\
  }\href {https://doi.org/10.1103/PhysRevLett.85.1092} {\bibfield  {journal}
  {\bibinfo  {journal} {Phys. Rev. Lett.}\ }\textbf {\bibinfo {volume} {85}},\
  \bibinfo {pages} {1092} (\bibinfo {year} {2000})}\BibitemShut {NoStop}%
\bibitem [{\citenamefont {Lux}\ \emph {et~al.}(2014)\citenamefont {Lux},
  \citenamefont {M{\"u}ller}, \citenamefont {Mitra},\ and\ \citenamefont
  {Rosch}}]{lux2014hydrodynamic}%
  \BibitemOpen
  \bibfield  {author} {\bibinfo {author} {\bibfnamefont {J.}~\bibnamefont
  {Lux}}, \bibinfo {author} {\bibfnamefont {J.}~\bibnamefont {M{\"u}ller}},
  \bibinfo {author} {\bibfnamefont {A.}~\bibnamefont {Mitra}},\ and\ \bibinfo
  {author} {\bibfnamefont {A.}~\bibnamefont {Rosch}},\ }\href@noop {}
  {\bibfield  {journal} {\bibinfo  {journal} {Physical Review A}\ }\textbf
  {\bibinfo {volume} {89}},\ \bibinfo {pages} {053608} (\bibinfo {year}
  {2014})}\BibitemShut {NoStop}%
\bibitem [{\citenamefont {Medenjak}\ \emph {et~al.}(2017)\citenamefont
  {Medenjak}, \citenamefont {Klobas},\ and\ \citenamefont
  {Prosen}}]{medenjak2017diffusion}%
  \BibitemOpen
  \bibfield  {author} {\bibinfo {author} {\bibfnamefont {M.}~\bibnamefont
  {Medenjak}}, \bibinfo {author} {\bibfnamefont {K.}~\bibnamefont {Klobas}},\
  and\ \bibinfo {author} {\bibfnamefont {T.}~\bibnamefont {Prosen}},\
  }\href@noop {} {\bibfield  {journal} {\bibinfo  {journal} {Physical review
  letters}\ }\textbf {\bibinfo {volume} {119}},\ \bibinfo {pages} {110603}
  (\bibinfo {year} {2017})}\BibitemShut {NoStop}%
\bibitem [{\citenamefont {Gopalakrishnan}\ and\ \citenamefont
  {Vasseur}(2019)}]{Gopalakrishnan2019}%
  \BibitemOpen
  \bibfield  {author} {\bibinfo {author} {\bibfnamefont {S.}~\bibnamefont
  {Gopalakrishnan}}\ and\ \bibinfo {author} {\bibfnamefont {R.}~\bibnamefont
  {Vasseur}},\ }\href {https://doi.org/10.1103/PhysRevLett.122.127202}
  {\bibfield  {journal} {\bibinfo  {journal} {Physical Review Letters}\
  }\textbf {\bibinfo {volume} {122}},\ \bibinfo {pages} {127202} (\bibinfo
  {year} {2019})}\BibitemShut {NoStop}%
\bibitem [{\citenamefont {Friedman}\ \emph {et~al.}(2020)\citenamefont
  {Friedman}, \citenamefont {Gopalakrishnan},\ and\ \citenamefont
  {Vasseur}}]{friedmanDiffusiveHydrodynamicsIntegrability2020}%
  \BibitemOpen
  \bibfield  {author} {\bibinfo {author} {\bibfnamefont {A.~J.}\ \bibnamefont
  {Friedman}}, \bibinfo {author} {\bibfnamefont {S.}~\bibnamefont
  {Gopalakrishnan}},\ and\ \bibinfo {author} {\bibfnamefont {R.}~\bibnamefont
  {Vasseur}},\ }\href {https://doi.org/10.1103/PhysRevB.101.180302} {\bibfield
  {journal} {\bibinfo  {journal} {Phys. Rev. B}\ }\textbf {\bibinfo {volume}
  {101}},\ \bibinfo {pages} {180302} (\bibinfo {year} {2020})}\BibitemShut
  {NoStop}%
\bibitem [{\citenamefont {Bertini}\ \emph {et~al.}(2021)\citenamefont
  {Bertini}, \citenamefont {Heidrich-Meisner}, \citenamefont {Karrasch},
  \citenamefont {Prosen}, \citenamefont {Steinigeweg},\ and\ \citenamefont
  {\ifmmode \check{Z}\else \v{Z}\fi{}nidari\ifmmode~\check{c}\else
  \v{c}\fi{}}}]{bertinifinite2021}%
  \BibitemOpen
  \bibfield  {author} {\bibinfo {author} {\bibfnamefont {B.}~\bibnamefont
  {Bertini}}, \bibinfo {author} {\bibfnamefont {F.}~\bibnamefont
  {Heidrich-Meisner}}, \bibinfo {author} {\bibfnamefont {C.}~\bibnamefont
  {Karrasch}}, \bibinfo {author} {\bibfnamefont {T.}~\bibnamefont {Prosen}},
  \bibinfo {author} {\bibfnamefont {R.}~\bibnamefont {Steinigeweg}},\ and\
  \bibinfo {author} {\bibfnamefont {M.}~\bibnamefont {\ifmmode \check{Z}\else
  \v{Z}\fi{}nidari\ifmmode~\check{c}\else \v{c}\fi{}}},\ }\href
  {https://doi.org/10.1103/RevModPhys.93.025003} {\bibfield  {journal}
  {\bibinfo  {journal} {Rev. Mod. Phys.}\ }\textbf {\bibinfo {volume} {93}},\
  \bibinfo {pages} {025003} (\bibinfo {year} {2021})}\BibitemShut {NoStop}%
\bibitem [{\citenamefont {Prosen}(2011{\natexlab{a}})}]{Prosen2011}%
  \BibitemOpen
  \bibfield  {author} {\bibinfo {author} {\bibfnamefont {T.}~\bibnamefont
  {Prosen}},\ }\href {https://doi.org/10.1103/PhysRevLett.106.217206}
  {\bibfield  {journal} {\bibinfo  {journal} {Physical Review Letters}\
  }\textbf {\bibinfo {volume} {106}},\ \bibinfo {pages} {2} (\bibinfo {year}
  {2011}{\natexlab{a}})}\BibitemShut {NoStop}%
\bibitem [{\citenamefont {Prosen}(2011{\natexlab{b}})}]{ProsenExactXXZ}%
  \BibitemOpen
  \bibfield  {author} {\bibinfo {author} {\bibfnamefont {T.}~\bibnamefont
  {Prosen}},\ }\href {https://doi.org/10.1103/PhysRevLett.107.137201}
  {\bibfield  {journal} {\bibinfo  {journal} {Phys. Rev. Lett.}\ }\textbf
  {\bibinfo {volume} {107}},\ \bibinfo {pages} {137201} (\bibinfo {year}
  {2011}{\natexlab{b}})}\BibitemShut {NoStop}%
\bibitem [{\citenamefont {Karevski}\ and\ \citenamefont
  {Platini}(2009)}]{KarevskiRepeatedinterac}%
  \BibitemOpen
  \bibfield  {author} {\bibinfo {author} {\bibfnamefont {D.}~\bibnamefont
  {Karevski}}\ and\ \bibinfo {author} {\bibfnamefont {T.}~\bibnamefont
  {Platini}},\ }\href {https://doi.org/10.1103/PhysRevLett.102.207207}
  {\bibfield  {journal} {\bibinfo  {journal} {Phys. Rev. Lett.}\ }\textbf
  {\bibinfo {volume} {102}},\ \bibinfo {pages} {207207} (\bibinfo {year}
  {2009})}\BibitemShut {NoStop}%
\bibitem [{\citenamefont {Karevski}\ \emph {et~al.}(2013)\citenamefont
  {Karevski}, \citenamefont {Popkov},\ and\ \citenamefont
  {Schütz}}]{KarevskiExactXXZ}%
  \BibitemOpen
  \bibfield  {author} {\bibinfo {author} {\bibfnamefont {D.}~\bibnamefont
  {Karevski}}, \bibinfo {author} {\bibfnamefont {V.}~\bibnamefont {Popkov}},\
  and\ \bibinfo {author} {\bibfnamefont {G.~M.}\ \bibnamefont {Schütz}},\
  }\href {https://doi.org/10.1103/PhysRevLett.110.047201} {\bibfield  {journal}
  {\bibinfo  {journal} {Phys. Rev. Lett.}\ }\textbf {\bibinfo {volume} {110}},\
  \bibinfo {pages} {047201} (\bibinfo {year} {2013})}\BibitemShut {NoStop}%
\bibitem [{\citenamefont {Ferreira}\ and\ \citenamefont
  {Filippone}(2020)}]{ferreiraBallistictodiffusiveTransitionSpin2020a}%
  \BibitemOpen
  \bibfield  {author} {\bibinfo {author} {\bibfnamefont {J.~S.}\ \bibnamefont
  {Ferreira}}\ and\ \bibinfo {author} {\bibfnamefont {M.}~\bibnamefont
  {Filippone}},\ }\href {https://doi.org/10.1103/PhysRevB.102.184304}
  {\bibfield  {journal} {\bibinfo  {journal} {Phys. Rev. B}\ }\textbf {\bibinfo
  {volume} {102}},\ \bibinfo {pages} {184304} (\bibinfo {year}
  {2020})}\BibitemShut {NoStop}%
\bibitem [{\citenamefont {Jin}\ \emph {et~al.}(2020{\natexlab{a}})\citenamefont
  {Jin}, \citenamefont {Filippone},\ and\ \citenamefont
  {Giamarchi}}]{JinFilipponeGiamarchi_GenericMarkovian}%
  \BibitemOpen
  \bibfield  {author} {\bibinfo {author} {\bibfnamefont {T.}~\bibnamefont
  {Jin}}, \bibinfo {author} {\bibfnamefont {M.}~\bibnamefont {Filippone}},\
  and\ \bibinfo {author} {\bibfnamefont {T.}~\bibnamefont {Giamarchi}},\ }\href
  {https://doi.org/10.1103/PhysRevB.102.205131} {\bibfield  {journal} {\bibinfo
   {journal} {Phys. Rev. B}\ }\textbf {\bibinfo {volume} {102}},\ \bibinfo
  {pages} {205131} (\bibinfo {year} {2020}{\natexlab{a}})}\BibitemShut
  {NoStop}%
\bibitem [{\citenamefont
  {{\v{Z}}nidari{\v{c}}}(2010{\natexlab{a}})}]{Znidaric__XXdeph}%
  \BibitemOpen
  \bibfield  {author} {\bibinfo {author} {\bibfnamefont {M.}~\bibnamefont
  {{\v{Z}}nidari{\v{c}}}},\ }\href
  {https://doi.org/10.1088/1742-5468/2010/05/l05002} {\bibfield  {journal}
  {\bibinfo  {journal} {Journal of Statistical Mechanics: Theory and
  Experiment}\ }\textbf {\bibinfo {volume} {2010}},\ \bibinfo {pages} {L05002}
  (\bibinfo {year} {2010}{\natexlab{a}})}\BibitemShut {NoStop}%
\bibitem [{\citenamefont {Bertini}\ \emph {et~al.}(2016)\citenamefont
  {Bertini}, \citenamefont {Collura}, \citenamefont {Nardis},\ and\
  \citenamefont {Fagotti}}]{Bertini_2016}%
  \BibitemOpen
  \bibfield  {author} {\bibinfo {author} {\bibfnamefont {B.}~\bibnamefont
  {Bertini}}, \bibinfo {author} {\bibfnamefont {M.}~\bibnamefont {Collura}},
  \bibinfo {author} {\bibfnamefont {J.~D.}\ \bibnamefont {Nardis}},\ and\
  \bibinfo {author} {\bibfnamefont {M.}~\bibnamefont {Fagotti}},\ }\href
  {https://doi.org/10.1103/physrevlett.117.207201} {\bibfield  {journal}
  {\bibinfo  {journal} {Physical Review Letters}\ }\textbf {\bibinfo {volume}
  {117}},\ \bibinfo {pages} {207201} (\bibinfo {year} {2016})}\BibitemShut
  {NoStop}%
\bibitem [{\citenamefont {M\"uller}\ \emph {et~al.}(2021)\citenamefont
  {M\"uller}, \citenamefont {Gievers}, \citenamefont {Fr\"oml}, \citenamefont
  {Diehl},\ and\ \citenamefont {Chiocchetta}}]{muller2021shape}%
  \BibitemOpen
  \bibfield  {author} {\bibinfo {author} {\bibfnamefont {T.}~\bibnamefont
  {M\"uller}}, \bibinfo {author} {\bibfnamefont {M.}~\bibnamefont {Gievers}},
  \bibinfo {author} {\bibfnamefont {H.}~\bibnamefont {Fr\"oml}}, \bibinfo
  {author} {\bibfnamefont {S.}~\bibnamefont {Diehl}},\ and\ \bibinfo {author}
  {\bibfnamefont {A.}~\bibnamefont {Chiocchetta}},\ }\href
  {https://doi.org/10.1103/PhysRevB.104.155431} {\bibfield  {journal} {\bibinfo
   {journal} {Phys. Rev. B}\ }\textbf {\bibinfo {volume} {104}},\ \bibinfo
  {pages} {155431} (\bibinfo {year} {2021})}\BibitemShut {NoStop}%
\bibitem [{\citenamefont {Rossini}\ \emph {et~al.}(2021)\citenamefont
  {Rossini}, \citenamefont {Ghermaoui}, \citenamefont {Aguilera}, \citenamefont
  {Vatr\'e}, \citenamefont {Bouganne}, \citenamefont {Beugnon}, \citenamefont
  {Gerbier},\ and\ \citenamefont {Mazza}}]{rossini2020strong}%
  \BibitemOpen
  \bibfield  {author} {\bibinfo {author} {\bibfnamefont {D.}~\bibnamefont
  {Rossini}}, \bibinfo {author} {\bibfnamefont {A.}~\bibnamefont {Ghermaoui}},
  \bibinfo {author} {\bibfnamefont {M.~B.}\ \bibnamefont {Aguilera}}, \bibinfo
  {author} {\bibfnamefont {R.}~\bibnamefont {Vatr\'e}}, \bibinfo {author}
  {\bibfnamefont {R.}~\bibnamefont {Bouganne}}, \bibinfo {author}
  {\bibfnamefont {J.}~\bibnamefont {Beugnon}}, \bibinfo {author} {\bibfnamefont
  {F.}~\bibnamefont {Gerbier}},\ and\ \bibinfo {author} {\bibfnamefont
  {L.}~\bibnamefont {Mazza}},\ }\href
  {https://doi.org/10.1103/PhysRevA.103.L060201} {\bibfield  {journal}
  {\bibinfo  {journal} {Phys. Rev. A}\ }\textbf {\bibinfo {volume} {103}},\
  \bibinfo {pages} {L060201} (\bibinfo {year} {2021})}\BibitemShut {NoStop}%
\bibitem [{\citenamefont {Alba}\ and\ \citenamefont
  {Carollo}(2022)}]{Alba_Losses}%
  \BibitemOpen
  \bibfield  {author} {\bibinfo {author} {\bibfnamefont {V.}~\bibnamefont
  {Alba}}\ and\ \bibinfo {author} {\bibfnamefont {F.}~\bibnamefont {Carollo}},\
  }\href {https://doi.org/10.1103/PhysRevB.105.054303} {\bibfield  {journal}
  {\bibinfo  {journal} {Phys. Rev. B}\ }\textbf {\bibinfo {volume} {105}},\
  \bibinfo {pages} {054303} (\bibinfo {year} {2022})}\BibitemShut {NoStop}%
\bibitem [{\citenamefont {{Visuri}}\ \emph {et~al.}(2022)\citenamefont
  {{Visuri}}, \citenamefont {{Giamarchi}},\ and\ \citenamefont
  {{Kollath}}}]{Visuri_Losses}%
  \BibitemOpen
  \bibfield  {author} {\bibinfo {author} {\bibfnamefont {A.~M.}\ \bibnamefont
  {{Visuri}}}, \bibinfo {author} {\bibfnamefont {T.}~\bibnamefont
  {{Giamarchi}}},\ and\ \bibinfo {author} {\bibfnamefont {C.}~\bibnamefont
  {{Kollath}}},\ }\href {https://arxiv.org/abs/2201.10286} {\bibfield
  {journal} {\bibinfo  {journal} {2201.10286}\ } (\bibinfo {year}
  {2022})}\BibitemShut {NoStop}%
\bibitem [{\citenamefont
  {{\v{Z}}nidari{\v{c}}}(2010{\natexlab{b}})}]{Znidaric__dephasingXXZ}%
  \BibitemOpen
  \bibfield  {author} {\bibinfo {author} {\bibfnamefont {M.}~\bibnamefont
  {{\v{Z}}nidari{\v{c}}}},\ }\href
  {https://doi.org/10.1088/1367-2630/12/4/043001} {\bibfield  {journal}
  {\bibinfo  {journal} {New Journal of Physics}\ }\textbf {\bibinfo {volume}
  {12}},\ \bibinfo {pages} {043001} (\bibinfo {year}
  {2010}{\natexlab{b}})}\BibitemShut {NoStop}%
\bibitem [{\citenamefont {Bastianello}\ \emph {et~al.}(2020)\citenamefont
  {Bastianello}, \citenamefont {De~Nardis},\ and\ \citenamefont
  {De~Luca}}]{BastianelloDeNardisDeluca_GHDDeph}%
  \BibitemOpen
  \bibfield  {author} {\bibinfo {author} {\bibfnamefont {A.}~\bibnamefont
  {Bastianello}}, \bibinfo {author} {\bibfnamefont {J.}~\bibnamefont
  {De~Nardis}},\ and\ \bibinfo {author} {\bibfnamefont {A.}~\bibnamefont
  {De~Luca}},\ }\href {https://doi.org/10.1103/PhysRevB.102.161110} {\bibfield
  {journal} {\bibinfo  {journal} {Phys. Rev. B}\ }\textbf {\bibinfo {volume}
  {102}},\ \bibinfo {pages} {161110} (\bibinfo {year} {2020})}\BibitemShut
  {NoStop}%
\bibitem [{\citenamefont
  {{Eisler}}(2011)}]{Eisler_CrossoverBallisticDiffusive}%
  \BibitemOpen
  \bibfield  {author} {\bibinfo {author} {\bibfnamefont {V.}~\bibnamefont
  {{Eisler}}},\ }\href {https://doi.org/10.1088/1742-5468/2011/06/P06007}
  {\bibfield  {journal} {\bibinfo  {journal} {Journal of Statistical Mechanics:
  Theory and Experiment}\ }\textbf {\bibinfo {volume} {2011}},\ \bibinfo
  {pages} {06007} (\bibinfo {year} {2011})}\BibitemShut {NoStop}%
\bibitem [{\citenamefont {Dolgirev}\ \emph {et~al.}(2020)\citenamefont
  {Dolgirev}, \citenamefont {Marino}, \citenamefont {Sels},\ and\ \citenamefont
  {Demler}}]{dolgirevNonGaussianCorrelationsImprinted2020a}%
  \BibitemOpen
  \bibfield  {author} {\bibinfo {author} {\bibfnamefont {P.~E.}\ \bibnamefont
  {Dolgirev}}, \bibinfo {author} {\bibfnamefont {J.}~\bibnamefont {Marino}},
  \bibinfo {author} {\bibfnamefont {D.}~\bibnamefont {Sels}},\ and\ \bibinfo
  {author} {\bibfnamefont {E.}~\bibnamefont {Demler}},\ }\href
  {https://doi.org/10.1103/PhysRevB.102.100301} {\bibfield  {journal} {\bibinfo
   {journal} {Phys. Rev. B}\ }\textbf {\bibinfo {volume} {102}},\ \bibinfo
  {pages} {100301} (\bibinfo {year} {2020})}\BibitemShut {NoStop}%
\bibitem [{\citenamefont {Wellnitz}\ \emph {et~al.}(2022)\citenamefont
  {Wellnitz}, \citenamefont {Preisser}, \citenamefont {Alba}, \citenamefont
  {Dubail},\ and\ \citenamefont {Schachenmayer}}]{Alba_EntanglementDephasing}%
  \BibitemOpen
  \bibfield  {author} {\bibinfo {author} {\bibfnamefont {D.}~\bibnamefont
  {Wellnitz}}, \bibinfo {author} {\bibfnamefont {G.}~\bibnamefont {Preisser}},
  \bibinfo {author} {\bibfnamefont {V.}~\bibnamefont {Alba}}, \bibinfo {author}
  {\bibfnamefont {J.}~\bibnamefont {Dubail}},\ and\ \bibinfo {author}
  {\bibfnamefont {J.}~\bibnamefont {Schachenmayer}},\ }\href
  {https://arxiv.org/abs/2201.05099} {\bibfield  {journal} {\bibinfo  {journal}
  {arXiv:2201.05099}\ } (\bibinfo {year} {2022})}\BibitemShut {NoStop}%
\bibitem [{\citenamefont {Bauer}\ \emph {et~al.}(2019)\citenamefont {Bauer},
  \citenamefont {Bernard},\ and\ \citenamefont
  {Jin}}]{BauerBernardJin_EquilibriumQSSEP}%
  \BibitemOpen
  \bibfield  {author} {\bibinfo {author} {\bibfnamefont {M.}~\bibnamefont
  {Bauer}}, \bibinfo {author} {\bibfnamefont {D.}~\bibnamefont {Bernard}},\
  and\ \bibinfo {author} {\bibfnamefont {T.}~\bibnamefont {Jin}},\ }\href
  {https://doi.org/10.21468/SciPostPhys.6.4.045} {\bibfield  {journal}
  {\bibinfo  {journal} {SciPost Phys.}\ }\textbf {\bibinfo {volume} {6}},\
  \bibinfo {pages} {45} (\bibinfo {year} {2019})}\BibitemShut {NoStop}%
\bibitem [{\citenamefont {Bernard}\ and\ \citenamefont
  {Jin}(2019)}]{BernardJin_QSSEP}%
  \BibitemOpen
  \bibfield  {author} {\bibinfo {author} {\bibfnamefont {D.}~\bibnamefont
  {Bernard}}\ and\ \bibinfo {author} {\bibfnamefont {T.}~\bibnamefont {Jin}},\
  }\href {https://doi.org/10.1103/PhysRevLett.123.080601} {\bibfield  {journal}
  {\bibinfo  {journal} {Phys. Rev. Lett.}\ }\textbf {\bibinfo {volume} {123}},\
  \bibinfo {pages} {080601} (\bibinfo {year} {2019})}\BibitemShut {NoStop}%
\bibitem [{\citenamefont {Jin}\ \emph {et~al.}(2020{\natexlab{b}})\citenamefont
  {Jin}, \citenamefont {Krajenbrink},\ and\ \citenamefont
  {Bernard}}]{JinKrajenbrinkBernard_QKPZ}%
  \BibitemOpen
  \bibfield  {author} {\bibinfo {author} {\bibfnamefont {T.}~\bibnamefont
  {Jin}}, \bibinfo {author} {\bibfnamefont {A.}~\bibnamefont {Krajenbrink}},\
  and\ \bibinfo {author} {\bibfnamefont {D.}~\bibnamefont {Bernard}},\ }\href
  {https://doi.org/10.1103/PhysRevLett.125.040603} {\bibfield  {journal}
  {\bibinfo  {journal} {Phys. Rev. Lett.}\ }\textbf {\bibinfo {volume} {125}},\
  \bibinfo {pages} {040603} (\bibinfo {year} {2020}{\natexlab{b}})}\BibitemShut
  {NoStop}%
\bibitem [{\citenamefont {Essler}\ and\ \citenamefont
  {Piroli}(2020)}]{EsslerPiroli_Operatorfragmentation}%
  \BibitemOpen
  \bibfield  {author} {\bibinfo {author} {\bibfnamefont {F.~H.~L.}\
  \bibnamefont {Essler}}\ and\ \bibinfo {author} {\bibfnamefont
  {L.}~\bibnamefont {Piroli}},\ }\href
  {https://doi.org/10.1103/PhysRevE.102.062210} {\bibfield  {journal} {\bibinfo
   {journal} {Phys. Rev. E}\ }\textbf {\bibinfo {volume} {102}},\ \bibinfo
  {pages} {062210} (\bibinfo {year} {2020})}\BibitemShut {NoStop}%
\bibitem [{\citenamefont {Bernard}\ and\ \citenamefont
  {Piroli}(2021)}]{BernardPiroli_QSSEPentanglement}%
  \BibitemOpen
  \bibfield  {author} {\bibinfo {author} {\bibfnamefont {D.}~\bibnamefont
  {Bernard}}\ and\ \bibinfo {author} {\bibfnamefont {L.}~\bibnamefont
  {Piroli}},\ }\href {https://doi.org/10.1103/PhysRevE.104.014146} {\bibfield
  {journal} {\bibinfo  {journal} {Phys. Rev. E}\ }\textbf {\bibinfo {volume}
  {104}},\ \bibinfo {pages} {014146} (\bibinfo {year} {2021})}\BibitemShut
  {NoStop}%
\bibitem [{\citenamefont {Bernard}\ and\ \citenamefont
  {Doussal}(2020)}]{BernardLeDoussal_StochasticCFT}%
  \BibitemOpen
  \bibfield  {author} {\bibinfo {author} {\bibfnamefont {D.}~\bibnamefont
  {Bernard}}\ and\ \bibinfo {author} {\bibfnamefont {P.~L.}\ \bibnamefont
  {Doussal}},\ }\href {https://doi.org/10.1209/0295-5075/131/10007} {\bibfield
  {journal} {\bibinfo  {journal} {{EPL} (Europhysics Letters)}\ }\textbf
  {\bibinfo {volume} {131}},\ \bibinfo {pages} {10007} (\bibinfo {year}
  {2020})}\BibitemShut {NoStop}%
\bibitem [{\citenamefont {Medvedyeva}\ \emph {et~al.}(2016)\citenamefont
  {Medvedyeva}, \citenamefont {Essler},\ and\ \citenamefont
  {Prosen}}]{ProsenEssler_Mapping}%
  \BibitemOpen
  \bibfield  {author} {\bibinfo {author} {\bibfnamefont {M.~V.}\ \bibnamefont
  {Medvedyeva}}, \bibinfo {author} {\bibfnamefont {F.~H.~L.}\ \bibnamefont
  {Essler}},\ and\ \bibinfo {author} {\bibfnamefont {T.}~\bibnamefont
  {Prosen}},\ }\href {https://doi.org/10.1103/PhysRevLett.117.137202}
  {\bibfield  {journal} {\bibinfo  {journal} {Phys. Rev. Lett.}\ }\textbf
  {\bibinfo {volume} {117}},\ \bibinfo {pages} {137202} (\bibinfo {year}
  {2016})}\BibitemShut {NoStop}%
\bibitem [{\citenamefont {Bauer}\ \emph {et~al.}(2017)\citenamefont {Bauer},
  \citenamefont {Bernard},\ and\ \citenamefont
  {Jin}}]{BauerBernardJin_Stoqdissipative}%
  \BibitemOpen
  \bibfield  {author} {\bibinfo {author} {\bibfnamefont {M.}~\bibnamefont
  {Bauer}}, \bibinfo {author} {\bibfnamefont {D.}~\bibnamefont {Bernard}},\
  and\ \bibinfo {author} {\bibfnamefont {T.}~\bibnamefont {Jin}},\ }\href
  {https://doi.org/10.21468/SciPostPhys.3.5.033} {\bibfield  {journal}
  {\bibinfo  {journal} {SciPost Phys.}\ }\textbf {\bibinfo {volume} {3}},\
  \bibinfo {pages} {033} (\bibinfo {year} {2017})}\BibitemShut {NoStop}%
\bibitem [{\citenamefont {Turkeshi}\ and\ \citenamefont
  {Schir\'o}(2021)}]{TurkeshiSchiro_Dephmodel}%
  \BibitemOpen
  \bibfield  {author} {\bibinfo {author} {\bibfnamefont {X.}~\bibnamefont
  {Turkeshi}}\ and\ \bibinfo {author} {\bibfnamefont {M.}~\bibnamefont
  {Schir\'o}},\ }\href {https://doi.org/10.1103/PhysRevB.104.144301} {\bibfield
   {journal} {\bibinfo  {journal} {Phys. Rev. B}\ }\textbf {\bibinfo {volume}
  {104}},\ \bibinfo {pages} {144301} (\bibinfo {year} {2021})}\BibitemShut
  {NoStop}%
\bibitem [{\citenamefont {Jin}\ \emph {et~al.}(2022)\citenamefont {Jin},
  \citenamefont {Ferreira}, \citenamefont {Filippone},\ and\ \citenamefont
  {Giamarchi}}]{Jin_Quantumresistors}%
  \BibitemOpen
  \bibfield  {author} {\bibinfo {author} {\bibfnamefont {T.}~\bibnamefont
  {Jin}}, \bibinfo {author} {\bibfnamefont {J.~S.}\ \bibnamefont {Ferreira}},
  \bibinfo {author} {\bibfnamefont {M.}~\bibnamefont {Filippone}},\ and\
  \bibinfo {author} {\bibfnamefont {T.}~\bibnamefont {Giamarchi}},\ }\href
  {https://doi.org/10.1103/PhysRevResearch.4.013109} {\bibfield  {journal}
  {\bibinfo  {journal} {Phys. Rev. Research}\ }\textbf {\bibinfo {volume}
  {4}},\ \bibinfo {pages} {013109} (\bibinfo {year} {2022})}\BibitemShut
  {NoStop}%
\bibitem [{\citenamefont {De~Nardis}\ \emph {et~al.}(2020)\citenamefont
  {De~Nardis}, \citenamefont {Gopalakrishnan}, \citenamefont {Ilievski},\ and\
  \citenamefont {Vasseur}}]{denardisSuperdiffusionEmergentClassical2020}%
  \BibitemOpen
  \bibfield  {author} {\bibinfo {author} {\bibfnamefont {J.}~\bibnamefont
  {De~Nardis}}, \bibinfo {author} {\bibfnamefont {S.}~\bibnamefont
  {Gopalakrishnan}}, \bibinfo {author} {\bibfnamefont {E.}~\bibnamefont
  {Ilievski}},\ and\ \bibinfo {author} {\bibfnamefont {R.}~\bibnamefont
  {Vasseur}},\ }\href {https://doi.org/10.1103/PhysRevLett.125.070601}
  {\bibfield  {journal} {\bibinfo  {journal} {Phys. Rev. Lett.}\ }\textbf
  {\bibinfo {volume} {125}},\ \bibinfo {pages} {070601} (\bibinfo {year}
  {2020})}\BibitemShut {NoStop}%
\bibitem [{\citenamefont {Zu}\ \emph {et~al.}(2021)\citenamefont {Zu},
  \citenamefont {Machado}, \citenamefont {Ye}, \citenamefont {Choi},
  \citenamefont {Kobrin}, \citenamefont {Mittiga}, \citenamefont {Hsieh},
  \citenamefont {Bhattacharyya}, \citenamefont {Markham}, \citenamefont
  {Twitchen}, \citenamefont {Jarmola}, \citenamefont {Budker}, \citenamefont
  {Laumann}, \citenamefont {Moore},\ and\ \citenamefont
  {Yao}}]{zuEmergentHydrodynamicsStrongly2021}%
  \BibitemOpen
  \bibfield  {author} {\bibinfo {author} {\bibfnamefont {C.}~\bibnamefont
  {Zu}}, \bibinfo {author} {\bibfnamefont {F.}~\bibnamefont {Machado}},
  \bibinfo {author} {\bibfnamefont {B.}~\bibnamefont {Ye}}, \bibinfo {author}
  {\bibfnamefont {S.}~\bibnamefont {Choi}}, \bibinfo {author} {\bibfnamefont
  {B.}~\bibnamefont {Kobrin}}, \bibinfo {author} {\bibfnamefont
  {T.}~\bibnamefont {Mittiga}}, \bibinfo {author} {\bibfnamefont
  {S.}~\bibnamefont {Hsieh}}, \bibinfo {author} {\bibfnamefont
  {P.}~\bibnamefont {Bhattacharyya}}, \bibinfo {author} {\bibfnamefont
  {M.}~\bibnamefont {Markham}}, \bibinfo {author} {\bibfnamefont
  {D.}~\bibnamefont {Twitchen}}, \bibinfo {author} {\bibfnamefont
  {A.}~\bibnamefont {Jarmola}}, \bibinfo {author} {\bibfnamefont
  {D.}~\bibnamefont {Budker}}, \bibinfo {author} {\bibfnamefont {C.~R.}\
  \bibnamefont {Laumann}}, \bibinfo {author} {\bibfnamefont {J.~E.}\
  \bibnamefont {Moore}},\ and\ \bibinfo {author} {\bibfnamefont {N.~Y.}\
  \bibnamefont {Yao}},\ }\href {https://doi.org/10.1038/s41586-021-03763-1}
  {\bibfield  {journal} {\bibinfo  {journal} {Nature}\ }\textbf {\bibinfo
  {volume} {597}},\ \bibinfo {pages} {45} (\bibinfo {year} {2021})}\BibitemShut
  {NoStop}%
\bibitem [{\citenamefont {Wurtz}\ \emph {et~al.}(2020)\citenamefont {Wurtz},
  \citenamefont {Claeys},\ and\ \citenamefont
  {Polkovnikov}}]{wurtz_variational_2020}%
  \BibitemOpen
  \bibfield  {author} {\bibinfo {author} {\bibfnamefont {J.}~\bibnamefont
  {Wurtz}}, \bibinfo {author} {\bibfnamefont {P.~W.}\ \bibnamefont {Claeys}},\
  and\ \bibinfo {author} {\bibfnamefont {A.}~\bibnamefont {Polkovnikov}},\
  }\href {https://doi.org/10.1103/PhysRevB.101.014302} {\bibfield  {journal}
  {\bibinfo  {journal} {Physical Review B}\ }\textbf {\bibinfo {volume}
  {101}},\ \bibinfo {pages} {014302} (\bibinfo {year} {2020})}\BibitemShut
  {NoStop}%
\bibitem [{\citenamefont {Nardis}\ \emph {et~al.}(2019)\citenamefont {Nardis},
  \citenamefont {Bernard},\ and\ \citenamefont
  {Doyon}}]{denardisDiffusionGeneralizedHydrodynamics2019}%
  \BibitemOpen
  \bibfield  {author} {\bibinfo {author} {\bibfnamefont {J.~D.}\ \bibnamefont
  {Nardis}}, \bibinfo {author} {\bibfnamefont {D.}~\bibnamefont {Bernard}},\
  and\ \bibinfo {author} {\bibfnamefont {B.}~\bibnamefont {Doyon}},\ }\href
  {https://doi.org/10.21468/SciPostPhys.6.4.049} {\bibfield  {journal}
  {\bibinfo  {journal} {SciPost Physics}\ }\textbf {\bibinfo {volume} {6}},\
  \bibinfo {pages} {049} (\bibinfo {year} {2019})}\BibitemShut {NoStop}%
\bibitem [{\citenamefont {Steinigeweg}\ \emph {et~al.}(2014)\citenamefont
  {Steinigeweg}, \citenamefont {Heidrich-Meisner}, \citenamefont {Gemmer},
  \citenamefont {Michielsen},\ and\ \citenamefont
  {Raedt}}]{steinigeweg2014scaling}%
  \BibitemOpen
  \bibfield  {author} {\bibinfo {author} {\bibfnamefont {R.}~\bibnamefont
  {Steinigeweg}}, \bibinfo {author} {\bibfnamefont {F.}~\bibnamefont
  {Heidrich-Meisner}}, \bibinfo {author} {\bibfnamefont {J.}~\bibnamefont
  {Gemmer}}, \bibinfo {author} {\bibfnamefont {K.}~\bibnamefont {Michielsen}},\
  and\ \bibinfo {author} {\bibfnamefont {H.~D.}\ \bibnamefont {Raedt}},\ }\href
  {https://doi.org/10.1103/physrevb.90.094417} {\bibfield  {journal} {\bibinfo
  {journal} {Physical Review B}\ }\textbf {\bibinfo {volume} {90}},\ \bibinfo
  {pages} {094417} (\bibinfo {year} {2014})}\BibitemShut {NoStop}%
\bibitem [{\citenamefont {Dogra}\ \emph {et~al.}(2019)\citenamefont {Dogra},
  \citenamefont {Landini}, \citenamefont {Kroeger}, \citenamefont {Hruby},
  \citenamefont {Donner},\ and\ \citenamefont
  {Esslinger}}]{dogra2019dissipation}%
  \BibitemOpen
  \bibfield  {author} {\bibinfo {author} {\bibfnamefont {N.}~\bibnamefont
  {Dogra}}, \bibinfo {author} {\bibfnamefont {M.}~\bibnamefont {Landini}},
  \bibinfo {author} {\bibfnamefont {K.}~\bibnamefont {Kroeger}}, \bibinfo
  {author} {\bibfnamefont {L.}~\bibnamefont {Hruby}}, \bibinfo {author}
  {\bibfnamefont {T.}~\bibnamefont {Donner}},\ and\ \bibinfo {author}
  {\bibfnamefont {T.}~\bibnamefont {Esslinger}},\ }\href@noop {} {\bibfield
  {journal} {\bibinfo  {journal} {Science}\ }\textbf {\bibinfo {volume}
  {366}},\ \bibinfo {pages} {1496} (\bibinfo {year} {2019})}\BibitemShut
  {NoStop}%
\bibitem [{\citenamefont {Ferri}\ \emph {et~al.}(2021)\citenamefont {Ferri},
  \citenamefont {Rosa-Medina}, \citenamefont {Finger}, \citenamefont {Dogra},
  \citenamefont {Soriente}, \citenamefont {Zilberberg}, \citenamefont
  {Donner},\ and\ \citenamefont {Esslinger}}]{ferri_2021}%
  \BibitemOpen
  \bibfield  {author} {\bibinfo {author} {\bibfnamefont {F.}~\bibnamefont
  {Ferri}}, \bibinfo {author} {\bibfnamefont {R.}~\bibnamefont {Rosa-Medina}},
  \bibinfo {author} {\bibfnamefont {F.}~\bibnamefont {Finger}}, \bibinfo
  {author} {\bibfnamefont {N.}~\bibnamefont {Dogra}}, \bibinfo {author}
  {\bibfnamefont {M.}~\bibnamefont {Soriente}}, \bibinfo {author}
  {\bibfnamefont {O.}~\bibnamefont {Zilberberg}}, \bibinfo {author}
  {\bibfnamefont {T.}~\bibnamefont {Donner}},\ and\ \bibinfo {author}
  {\bibfnamefont {T.}~\bibnamefont {Esslinger}},\ }\href
  {https://doi.org/10.1103/PhysRevX.11.041046} {\bibfield  {journal} {\bibinfo
  {journal} {Phys. Rev. X}\ }\textbf {\bibinfo {volume} {11}},\ \bibinfo
  {pages} {041046} (\bibinfo {year} {2021})}\BibitemShut {NoStop}%
\bibitem [{\citenamefont {Rosa-Medina}\ \emph {et~al.}(2022)\citenamefont
  {Rosa-Medina}, \citenamefont {Ferri}, \citenamefont {Finger}, \citenamefont
  {Dogra}, \citenamefont {Kroeger}, \citenamefont {Lin}, \citenamefont
  {Chitra}, \citenamefont {Donner},\ and\ \citenamefont
  {Esslinger}}]{rosamedina_2022}%
  \BibitemOpen
  \bibfield  {author} {\bibinfo {author} {\bibfnamefont {R.}~\bibnamefont
  {Rosa-Medina}}, \bibinfo {author} {\bibfnamefont {F.}~\bibnamefont {Ferri}},
  \bibinfo {author} {\bibfnamefont {F.}~\bibnamefont {Finger}}, \bibinfo
  {author} {\bibfnamefont {N.}~\bibnamefont {Dogra}}, \bibinfo {author}
  {\bibfnamefont {K.}~\bibnamefont {Kroeger}}, \bibinfo {author} {\bibfnamefont
  {R.}~\bibnamefont {Lin}}, \bibinfo {author} {\bibfnamefont {R.}~\bibnamefont
  {Chitra}}, \bibinfo {author} {\bibfnamefont {T.}~\bibnamefont {Donner}},\
  and\ \bibinfo {author} {\bibfnamefont {T.}~\bibnamefont {Esslinger}},\ }\href
  {https://doi.org/10.1103/PhysRevLett.128.143602} {\bibfield  {journal}
  {\bibinfo  {journal} {Phys. Rev. Lett.}\ }\textbf {\bibinfo {volume} {128}},\
  \bibinfo {pages} {143602} (\bibinfo {year} {2022})}\BibitemShut {NoStop}%
\bibitem [{\citenamefont {Corman}\ \emph {et~al.}(2019)\citenamefont {Corman},
  \citenamefont {Fabritius}, \citenamefont {H\"ausler}, \citenamefont {Mohan},
  \citenamefont {Dogra}, \citenamefont {Husmann}, \citenamefont {Lebrat},\ and\
  \citenamefont {Esslinger}}]{corman_2019}%
  \BibitemOpen
  \bibfield  {author} {\bibinfo {author} {\bibfnamefont {L.}~\bibnamefont
  {Corman}}, \bibinfo {author} {\bibfnamefont {P.}~\bibnamefont {Fabritius}},
  \bibinfo {author} {\bibfnamefont {S.}~\bibnamefont {H\"ausler}}, \bibinfo
  {author} {\bibfnamefont {J.}~\bibnamefont {Mohan}}, \bibinfo {author}
  {\bibfnamefont {L.~H.}\ \bibnamefont {Dogra}}, \bibinfo {author}
  {\bibfnamefont {D.}~\bibnamefont {Husmann}}, \bibinfo {author} {\bibfnamefont
  {M.}~\bibnamefont {Lebrat}},\ and\ \bibinfo {author} {\bibfnamefont
  {T.}~\bibnamefont {Esslinger}},\ }\href
  {https://doi.org/10.1103/PhysRevA.100.053605} {\bibfield  {journal} {\bibinfo
   {journal} {Phys. Rev. A}\ }\textbf {\bibinfo {volume} {100}},\ \bibinfo
  {pages} {053605} (\bibinfo {year} {2019})}\BibitemShut {NoStop}%
\bibitem [{\citenamefont {Lebrat}\ \emph {et~al.}(2019)\citenamefont {Lebrat},
  \citenamefont {H\"ausler}, \citenamefont {Fabritius}, \citenamefont
  {Husmann}, \citenamefont {Corman},\ and\ \citenamefont
  {Esslinger}}]{lebratquantized2019}%
  \BibitemOpen
  \bibfield  {author} {\bibinfo {author} {\bibfnamefont {M.}~\bibnamefont
  {Lebrat}}, \bibinfo {author} {\bibfnamefont {S.}~\bibnamefont {H\"ausler}},
  \bibinfo {author} {\bibfnamefont {P.}~\bibnamefont {Fabritius}}, \bibinfo
  {author} {\bibfnamefont {D.}~\bibnamefont {Husmann}}, \bibinfo {author}
  {\bibfnamefont {L.}~\bibnamefont {Corman}},\ and\ \bibinfo {author}
  {\bibfnamefont {T.}~\bibnamefont {Esslinger}},\ }\href
  {https://doi.org/10.1103/PhysRevLett.123.193605} {\bibfield  {journal}
  {\bibinfo  {journal} {Phys. Rev. Lett.}\ }\textbf {\bibinfo {volume} {123}},\
  \bibinfo {pages} {193605} (\bibinfo {year} {2019})}\BibitemShut {NoStop}%
\bibitem [{\citenamefont {Mancini}\ \emph {et~al.}(2015)\citenamefont
  {Mancini}, \citenamefont {Pagano}, \citenamefont {Cappellini}, \citenamefont
  {Livi}, \citenamefont {Rider}, \citenamefont {Catani}, \citenamefont {Sias},
  \citenamefont {Zoller}, \citenamefont {Inguscio}, \citenamefont {Dalmonte},\
  and\ \citenamefont {Fallani}}]{Mancini1510}%
  \BibitemOpen
  \bibfield  {author} {\bibinfo {author} {\bibfnamefont {M.}~\bibnamefont
  {Mancini}}, \bibinfo {author} {\bibfnamefont {G.}~\bibnamefont {Pagano}},
  \bibinfo {author} {\bibfnamefont {G.}~\bibnamefont {Cappellini}}, \bibinfo
  {author} {\bibfnamefont {L.}~\bibnamefont {Livi}}, \bibinfo {author}
  {\bibfnamefont {M.}~\bibnamefont {Rider}}, \bibinfo {author} {\bibfnamefont
  {J.}~\bibnamefont {Catani}}, \bibinfo {author} {\bibfnamefont
  {C.}~\bibnamefont {Sias}}, \bibinfo {author} {\bibfnamefont {P.}~\bibnamefont
  {Zoller}}, \bibinfo {author} {\bibfnamefont {M.}~\bibnamefont {Inguscio}},
  \bibinfo {author} {\bibfnamefont {M.}~\bibnamefont {Dalmonte}},\ and\
  \bibinfo {author} {\bibfnamefont {L.}~\bibnamefont {Fallani}},\ }\href
  {https://doi.org/10.1126/science.aaa8736} {\bibfield  {journal} {\bibinfo
  {journal} {Science}\ }\textbf {\bibinfo {volume} {349}},\ \bibinfo {pages}
  {1510} (\bibinfo {year} {2015})}\BibitemShut {NoStop}%
\bibitem [{\citenamefont {Genkina}\ \emph {et~al.}(2019)\citenamefont
  {Genkina}, \citenamefont {Aycock}, \citenamefont {Lu}, \citenamefont {Lu},
  \citenamefont {Pineiro},\ and\ \citenamefont
  {Spielman}}]{genkina2019imaging}%
  \BibitemOpen
  \bibfield  {author} {\bibinfo {author} {\bibfnamefont {D.}~\bibnamefont
  {Genkina}}, \bibinfo {author} {\bibfnamefont {L.~M.}\ \bibnamefont {Aycock}},
  \bibinfo {author} {\bibfnamefont {H.-I.}\ \bibnamefont {Lu}}, \bibinfo
  {author} {\bibfnamefont {M.}~\bibnamefont {Lu}}, \bibinfo {author}
  {\bibfnamefont {A.~M.}\ \bibnamefont {Pineiro}},\ and\ \bibinfo {author}
  {\bibfnamefont {I.}~\bibnamefont {Spielman}},\ }\href
  {https://iopscience.iop.org/article/10.1088/1367-2630/ab165b} {\bibfield
  {journal} {\bibinfo  {journal} {New journal of physics}\ }\textbf {\bibinfo
  {volume} {21}},\ \bibinfo {pages} {053021} (\bibinfo {year}
  {2019})}\BibitemShut {NoStop}%
\bibitem [{\citenamefont {Chalopin}\ \emph {et~al.}(2020)\citenamefont
  {Chalopin}, \citenamefont {Satoor}, \citenamefont {Evrard}, \citenamefont
  {Makhalov}, \citenamefont {Dalibard}, \citenamefont {Lopes},\ and\
  \citenamefont {Nascimbene}}]{chalopin_probing_2020}%
  \BibitemOpen
  \bibfield  {author} {\bibinfo {author} {\bibfnamefont {T.}~\bibnamefont
  {Chalopin}}, \bibinfo {author} {\bibfnamefont {T.}~\bibnamefont {Satoor}},
  \bibinfo {author} {\bibfnamefont {A.}~\bibnamefont {Evrard}}, \bibinfo
  {author} {\bibfnamefont {V.}~\bibnamefont {Makhalov}}, \bibinfo {author}
  {\bibfnamefont {J.}~\bibnamefont {Dalibard}}, \bibinfo {author}
  {\bibfnamefont {R.}~\bibnamefont {Lopes}},\ and\ \bibinfo {author}
  {\bibfnamefont {S.}~\bibnamefont {Nascimbene}},\ }\href
  {https://doi.org/10.1038/s41567-020-0942-5} {\bibfield  {journal} {\bibinfo
  {journal} {Nature Physics}\ }\textbf {\bibinfo {volume} {16}},\ \bibinfo
  {pages} {1017} (\bibinfo {year} {2020})}\BibitemShut {NoStop}%
\bibitem [{\citenamefont {Zhou}\ \emph {et~al.}(2022)\citenamefont {Zhou},
  \citenamefont {Cappellini}, \citenamefont {Tusi}, \citenamefont {Franchi},
  \citenamefont {Parravicini}, \citenamefont {Repellin}, \citenamefont
  {Greschner}, \citenamefont {Inguscio}, \citenamefont {Giamarchi},
  \citenamefont {Filippone}, \citenamefont {Catani},\ and\ \citenamefont
  {Fallani}}]{zhou2022observation}%
  \BibitemOpen
  \bibfield  {author} {\bibinfo {author} {\bibfnamefont {T.~W.}\ \bibnamefont
  {Zhou}}, \bibinfo {author} {\bibfnamefont {G.}~\bibnamefont {Cappellini}},
  \bibinfo {author} {\bibfnamefont {D.}~\bibnamefont {Tusi}}, \bibinfo {author}
  {\bibfnamefont {L.}~\bibnamefont {Franchi}}, \bibinfo {author} {\bibfnamefont
  {J.}~\bibnamefont {Parravicini}}, \bibinfo {author} {\bibfnamefont
  {C.}~\bibnamefont {Repellin}}, \bibinfo {author} {\bibfnamefont
  {S.}~\bibnamefont {Greschner}}, \bibinfo {author} {\bibfnamefont
  {M.}~\bibnamefont {Inguscio}}, \bibinfo {author} {\bibfnamefont
  {T.}~\bibnamefont {Giamarchi}}, \bibinfo {author} {\bibfnamefont
  {M.}~\bibnamefont {Filippone}}, \bibinfo {author} {\bibfnamefont
  {J.}~\bibnamefont {Catani}},\ and\ \bibinfo {author} {\bibfnamefont
  {L.}~\bibnamefont {Fallani}},\ }\href {https://arxiv.org/abs/2205.13567}
  {\bibfield  {journal} {\bibinfo  {journal} {arXiv:2205.13567}\ } (\bibinfo
  {year} {2022})}\BibitemShut {NoStop}%
\bibitem [{\citenamefont {{\O}ksendal}(2003)}]{Oksendal_book}%
  \BibitemOpen
  \bibfield  {author} {\bibinfo {author} {\bibfnamefont {B.}~\bibnamefont
  {{\O}ksendal}},\ }\href {https://doi.org/10.1007/978-3-642-14394-6} {\emph
  {\bibinfo {title} {Stochastic Differential Equations}}}\ (\bibinfo
  {publisher} {Springer Berlin Heidelberg},\ \bibinfo {year}
  {2003})\BibitemShut {NoStop}%
\bibitem [{\citenamefont {Kamenev}(2011)}]{Kamenev2011}%
  \BibitemOpen
  \bibfield  {author} {\bibinfo {author} {\bibfnamefont {A.}~\bibnamefont
  {Kamenev}},\ }\href {https://doi.org/10.1017/CBO9781139003667} {\emph
  {\bibinfo {title} {Field Theory of Non-Equilibrium Systems}}}\ (\bibinfo
  {publisher} {Cambridge University Press},\ \bibinfo {address} {Cambridge},\
  \bibinfo {year} {2011})\ pp.\ \bibinfo {pages} {1--341}\BibitemShut {NoStop}%
\bibitem [{Note1()}]{Note1}%
  \BibitemOpen
  \bibinfo {note} {The Green's functions depend on the time differences $t-t'$,
  instead of separate times $t,t'$, as we consider situations where both the
  Hamiltonian~\protect \textup {\hbox {\mathsurround \z@ \protect \normalfont
  (\ignorespaces \ref {eq:system}\unskip \@@italiccorr )}} and Lindblad
  generator~\protect \textup {\hbox {\mathsurround \z@ \protect \normalfont
  (\ignorespaces \ref {eq:lindblad}\unskip \@@italiccorr )}} do not depend
  explicitly on time.}\BibitemShut {Stop}%
\bibitem [{\citenamefont {Larkin}\ and\ \citenamefont
  {Ovchinnikov}(1977)}]{Larkin_vortices_supra}%
  \BibitemOpen
  \bibfield  {author} {\bibinfo {author} {\bibfnamefont {A.~I.}\ \bibnamefont
  {Larkin}}\ and\ \bibinfo {author} {\bibfnamefont {I.~N.}\ \bibnamefont
  {Ovchinnikov}},\ }\href
  {https://ui.adsabs.harvard.edu/abs/1977ZhETF..73..299L/abstract} {\bibfield
  {journal} {\bibinfo  {journal} {Zhurnal Eksperimentalnoi i Teoreticheskoi
  Fiziki}\ }\textbf {\bibinfo {volume} {73}},\ \bibinfo {pages} {299} (\bibinfo
  {year} {1977})}\BibitemShut {NoStop}%
\bibitem [{\citenamefont {Meir}\ and\ \citenamefont
  {Wingreen}(1992)}]{MeirWingreenformula}%
  \BibitemOpen
  \bibfield  {author} {\bibinfo {author} {\bibfnamefont {Y.}~\bibnamefont
  {Meir}}\ and\ \bibinfo {author} {\bibfnamefont {N.~S.}\ \bibnamefont
  {Wingreen}},\ }\href {https://doi.org/10.1103/PhysRevLett.68.2512} {\bibfield
   {journal} {\bibinfo  {journal} {Phys. Rev. Lett.}\ }\textbf {\bibinfo
  {volume} {68}},\ \bibinfo {pages} {2512} (\bibinfo {year}
  {1992})}\BibitemShut {NoStop}%
\bibitem [{\citenamefont {Landauer}(1957)}]{LandauerFormula}%
  \BibitemOpen
  \bibfield  {author} {\bibinfo {author} {\bibfnamefont {R.}~\bibnamefont
  {Landauer}},\ }\href {https://doi.org/10.1147/rd.13.0223} {\bibfield
  {journal} {\bibinfo  {journal} {IBM Journal of Research and Development}\
  }\textbf {\bibinfo {volume} {1}},\ \bibinfo {pages} {223} (\bibinfo {year}
  {1957})}\BibitemShut {NoStop}%
\bibitem [{\citenamefont {Datta}(1997)}]{datta_electronic_1997}%
  \BibitemOpen
  \bibfield  {author} {\bibinfo {author} {\bibfnamefont {S.}~\bibnamefont
  {Datta}},\ }\href {https://doi.org/10.2139/ssrn.1029228} {\emph {\bibinfo
  {title} {Electronic transport in mesoscopic systems}}}\ (\bibinfo
  {publisher} {Cambridge university press},\ \bibinfo {year}
  {1997})\BibitemShut {NoStop}%
\bibitem [{\citenamefont {Lesovik}\ and\ \citenamefont
  {Sadovskyy}(2011)}]{Lesovik_2011}%
  \BibitemOpen
  \bibfield  {author} {\bibinfo {author} {\bibfnamefont {G.~B.}\ \bibnamefont
  {Lesovik}}\ and\ \bibinfo {author} {\bibfnamefont {I.~A.}\ \bibnamefont
  {Sadovskyy}},\ }\href {https://doi.org/10.3367/ufne.0181.201110b.1041}
  {\bibfield  {journal} {\bibinfo  {journal} {Physics-Uspekhi}\ }\textbf
  {\bibinfo {volume} {54}},\ \bibinfo {pages} {1007} (\bibinfo {year}
  {2011})}\BibitemShut {NoStop}%
\bibitem [{\citenamefont {Nazarov}\ and\ \citenamefont
  {Blanter}(2009)}]{nazarov_quantum_2009}%
  \BibitemOpen
  \bibfield  {author} {\bibinfo {author} {\bibfnamefont {Y.~V.}\ \bibnamefont
  {Nazarov}}\ and\ \bibinfo {author} {\bibfnamefont {Y.~M.}\ \bibnamefont
  {Blanter}},\ }\href {https://doi.org/10.1017/CBO9780511626906} {\emph
  {\bibinfo {title} {Quantum {Transport}: {Introduction} to {Nanoscience}}}}\
  (\bibinfo  {publisher} {Cambridge University Press},\ \bibinfo {address}
  {Cambridge},\ \bibinfo {year} {2009})\BibitemShut {NoStop}%
\bibitem [{\citenamefont {{\v{Z}}nidari{\v{c}}}\ and\ \citenamefont
  {Horvat}(2013)}]{Znidaric_dephasing}%
  \BibitemOpen
  \bibfield  {author} {\bibinfo {author} {\bibfnamefont {M.}~\bibnamefont
  {{\v{Z}}nidari{\v{c}}}}\ and\ \bibinfo {author} {\bibfnamefont
  {M.}~\bibnamefont {Horvat}},\ }\href
  {https://doi.org/10.1140/epjb/e2012-30730-9} {\bibfield  {journal} {\bibinfo
  {journal} {The European Physical Journal B}\ }\textbf {\bibinfo {volume}
  {86}},\ \bibinfo {pages} {67} (\bibinfo {year} {2013})}\BibitemShut {NoStop}%
\bibitem [{\citenamefont {Cai}\ and\ \citenamefont
  {Barthel}(2013)}]{CaiZi_AlgebraicvsExponentialDissipativesystems}%
  \BibitemOpen
  \bibfield  {author} {\bibinfo {author} {\bibfnamefont {Z.}~\bibnamefont
  {Cai}}\ and\ \bibinfo {author} {\bibfnamefont {T.}~\bibnamefont {Barthel}},\
  }\href {https://doi.org/10.1103/PhysRevLett.111.150403} {\bibfield  {journal}
  {\bibinfo  {journal} {Phys. Rev. Lett.}\ }\textbf {\bibinfo {volume} {111}},\
  \bibinfo {pages} {150403} (\bibinfo {year} {2013})}\BibitemShut {NoStop}%
\bibitem [{Note2()}]{Note2}%
  \BibitemOpen
  \bibinfo {note} {Despite the possibility to rely on Eq.~\protect \textup
  {\hbox {\mathsurround \z@ \protect \normalfont (\ignorespaces \ref
  {eq:LB}\unskip \@@italiccorr )}} to calculate the conductance for $\gamma
  >0$, we found more practical to perform the direct numerical calculation of
  the current as expressed in Eq.~\protect \textup {\hbox {\mathsurround \z@
  \protect \normalfont (\ignorespaces \ref {eq:current}\unskip \@@italiccorr
  )}} directly in the linear regime to derive the conductance~\protect \textup
  {\hbox {\mathsurround \z@ \protect \normalfont (\ignorespaces \ref
  {eq:conductance}\unskip \@@italiccorr )}}}\BibitemShut {NoStop}%
\bibitem [{Note3()}]{Note3}%
  \BibitemOpen
  \bibinfo {note} {The different scalings of $D$ and $G$ with $\mu ,T$ indicate
  that the contact resistance between bulk and leads is extensive with the
  system size. Fig.\ref {fig:heat} and previous works~\protect \citep
  {TurkeshiSchiro_Dephmodel} suggest that thermalization only occurs very deep
  in the bulk, supporting this hypothesis.}\BibitemShut {Stop}%
\bibitem [{\citenamefont {Dalibard}\ \emph {et~al.}(1992)\citenamefont
  {Dalibard}, \citenamefont {Castin},\ and\ \citenamefont
  {M{\o}lmer}}]{Dalibard_Unraveling}%
  \BibitemOpen
  \bibfield  {author} {\bibinfo {author} {\bibfnamefont {J.}~\bibnamefont
  {Dalibard}}, \bibinfo {author} {\bibfnamefont {Y.}~\bibnamefont {Castin}},\
  and\ \bibinfo {author} {\bibfnamefont {K.}~\bibnamefont {M{\o}lmer}},\ }\href
  {https://doi.org/10.1103/PhysRevLett.68.580} {\bibfield  {journal} {\bibinfo
  {journal} {Phys. Rev. Lett.}\ }\textbf {\bibinfo {volume} {68}},\ \bibinfo
  {pages} {580} (\bibinfo {year} {1992})}\BibitemShut {NoStop}%
\bibitem [{\citenamefont {Belavkin}(1990)}]{Belavkin_1990}%
  \BibitemOpen
  \bibfield  {author} {\bibinfo {author} {\bibfnamefont {V.}~\bibnamefont
  {Belavkin}},\ }\href {https://doi.org/10.1007/bfb0085504} {\bibfield
  {journal} {\bibinfo  {journal} {Journal of Soviet Mathematics v}\ ,\ \bibinfo
  {pages} {99}} (\bibinfo {year} {1990})}\BibitemShut {NoStop}%
\bibitem [{Note4()}]{Note4}%
  \BibitemOpen
  \bibinfo {note} {Convergence is exponential with the number of iterations and
  independent of the initial guess for $P(x)$, which we take
  arbitrarily.}\BibitemShut {Stop}%
\bibitem [{\citenamefont {Cao}\ \emph {et~al.}(2019)\citenamefont {Cao},
  \citenamefont {Tilloy},\ and\ \citenamefont {Luca}}]{Cao_2019}%
  \BibitemOpen
  \bibfield  {author} {\bibinfo {author} {\bibfnamefont {X.}~\bibnamefont
  {Cao}}, \bibinfo {author} {\bibfnamefont {A.}~\bibnamefont {Tilloy}},\ and\
  \bibinfo {author} {\bibfnamefont {A.~D.}\ \bibnamefont {Luca}},\ }\href
  {https://scipost.org/10.21468/SciPostPhys.7.2.024} {\bibfield  {journal}
  {\bibinfo  {journal} {{SciPost} Physics}\ }\textbf {\bibinfo {volume} {7}}
  (\bibinfo {year} {2019})}\BibitemShut {NoStop}%
\bibitem [{\citenamefont {Turkeshi}\ \emph {et~al.}(2021)\citenamefont
  {Turkeshi}, \citenamefont {Dalmonte}, \citenamefont {Fazio},\ and\
  \citenamefont {Schirò}}]{Turkeshi2021}%
  \BibitemOpen
  \bibfield  {author} {\bibinfo {author} {\bibfnamefont {X.}~\bibnamefont
  {Turkeshi}}, \bibinfo {author} {\bibfnamefont {M.}~\bibnamefont {Dalmonte}},
  \bibinfo {author} {\bibfnamefont {R.}~\bibnamefont {Fazio}},\ and\ \bibinfo
  {author} {\bibfnamefont {M.}~\bibnamefont {Schirò}},\ }\href
  {https://arxiv.org/pdf/2111.03500.pdf} {\bibfield  {journal} {\bibinfo
  {journal} {arXiv:2111.03500}\ } (\bibinfo {year} {2021})}\BibitemShut
  {NoStop}%
\bibitem [{\citenamefont {Ozawa}\ \emph {et~al.}(2016)\citenamefont {Ozawa},
  \citenamefont {Price}, \citenamefont {Goldman}, \citenamefont {Zilberberg},\
  and\ \citenamefont {Carusotto}}]{ozawa2016synthetic}%
  \BibitemOpen
  \bibfield  {author} {\bibinfo {author} {\bibfnamefont {T.}~\bibnamefont
  {Ozawa}}, \bibinfo {author} {\bibfnamefont {H.~M.}\ \bibnamefont {Price}},
  \bibinfo {author} {\bibfnamefont {N.}~\bibnamefont {Goldman}}, \bibinfo
  {author} {\bibfnamefont {O.}~\bibnamefont {Zilberberg}},\ and\ \bibinfo
  {author} {\bibfnamefont {I.}~\bibnamefont {Carusotto}},\ }\href
  {https://doi.org/10.1103/physreva.93.043827} {\bibfield  {journal} {\bibinfo
  {journal} {Physical Review A}\ }\textbf {\bibinfo {volume} {93}},\ \bibinfo
  {pages} {043827} (\bibinfo {year} {2016})}\BibitemShut {NoStop}%
\bibitem [{\citenamefont {Mittal}\ \emph {et~al.}(2019)\citenamefont {Mittal},
  \citenamefont {Orre}, \citenamefont {Leykam}, \citenamefont {Chong},\ and\
  \citenamefont {Hafezi}}]{mittal2019photonic}%
  \BibitemOpen
  \bibfield  {author} {\bibinfo {author} {\bibfnamefont {S.}~\bibnamefont
  {Mittal}}, \bibinfo {author} {\bibfnamefont {V.~V.}\ \bibnamefont {Orre}},
  \bibinfo {author} {\bibfnamefont {D.}~\bibnamefont {Leykam}}, \bibinfo
  {author} {\bibfnamefont {Y.~D.}\ \bibnamefont {Chong}},\ and\ \bibinfo
  {author} {\bibfnamefont {M.}~\bibnamefont {Hafezi}},\ }\href@noop {}
  {\bibfield  {journal} {\bibinfo  {journal} {Physical review letters}\
  }\textbf {\bibinfo {volume} {123}},\ \bibinfo {pages} {043201} (\bibinfo
  {year} {2019})}\BibitemShut {NoStop}%
\bibitem [{\citenamefont {Salerno}\ \emph {et~al.}(2019)\citenamefont
  {Salerno}, \citenamefont {Price}, \citenamefont {Lebrat}, \citenamefont
  {H{\"a}usler}, \citenamefont {Esslinger}, \citenamefont {Corman},
  \citenamefont {Brantut},\ and\ \citenamefont
  {Goldman}}]{salerno_quantized_2019}%
  \BibitemOpen
  \bibfield  {author} {\bibinfo {author} {\bibfnamefont {G.}~\bibnamefont
  {Salerno}}, \bibinfo {author} {\bibfnamefont {H.~M.}\ \bibnamefont {Price}},
  \bibinfo {author} {\bibfnamefont {M.}~\bibnamefont {Lebrat}}, \bibinfo
  {author} {\bibfnamefont {S.}~\bibnamefont {H{\"a}usler}}, \bibinfo {author}
  {\bibfnamefont {T.}~\bibnamefont {Esslinger}}, \bibinfo {author}
  {\bibfnamefont {L.}~\bibnamefont {Corman}}, \bibinfo {author} {\bibfnamefont
  {J.-P.}\ \bibnamefont {Brantut}},\ and\ \bibinfo {author} {\bibfnamefont
  {N.}~\bibnamefont {Goldman}},\ }\href
  {https://doi.org/10.1103/PhysRevX.9.041001} {\bibfield  {journal} {\bibinfo
  {journal} {Physical Review X}\ }\textbf {\bibinfo {volume} {9}},\ \bibinfo
  {pages} {041001} (\bibinfo {year} {2019})}\BibitemShut {NoStop}%
\bibitem [{\citenamefont {Livi}\ \emph {et~al.}(2016)\citenamefont {Livi},
  \citenamefont {Cappellini}, \citenamefont {Diem}, \citenamefont {Franchi},
  \citenamefont {Clivati}, \citenamefont {Frittelli}, \citenamefont {Levi},
  \citenamefont {Calonico}, \citenamefont {Catani}, \citenamefont {Inguscio}
  \emph {et~al.}}]{livi2016synthetic}%
  \BibitemOpen
  \bibfield  {author} {\bibinfo {author} {\bibfnamefont {L.}~\bibnamefont
  {Livi}}, \bibinfo {author} {\bibfnamefont {G.}~\bibnamefont {Cappellini}},
  \bibinfo {author} {\bibfnamefont {M.}~\bibnamefont {Diem}}, \bibinfo {author}
  {\bibfnamefont {L.}~\bibnamefont {Franchi}}, \bibinfo {author} {\bibfnamefont
  {C.}~\bibnamefont {Clivati}}, \bibinfo {author} {\bibfnamefont
  {M.}~\bibnamefont {Frittelli}}, \bibinfo {author} {\bibfnamefont
  {F.}~\bibnamefont {Levi}}, \bibinfo {author} {\bibfnamefont {D.}~\bibnamefont
  {Calonico}}, \bibinfo {author} {\bibfnamefont {J.}~\bibnamefont {Catani}},
  \bibinfo {author} {\bibfnamefont {M.}~\bibnamefont {Inguscio}}, \emph
  {et~al.},\ }\href@noop {} {\bibfield  {journal} {\bibinfo  {journal}
  {Physical review letters}\ }\textbf {\bibinfo {volume} {117}},\ \bibinfo
  {pages} {220401} (\bibinfo {year} {2016})}\BibitemShut {NoStop}%
\bibitem [{\citenamefont {Usmani}(1994)}]{usmani1994inversion}%
  \BibitemOpen
  \bibfield  {author} {\bibinfo {author} {\bibfnamefont {R.~A.}\ \bibnamefont
  {Usmani}},\ }\href
  {https://doi.org/https://doi.org/10.1016/0024-3795(94)90414-6} {\bibfield
  {journal} {\bibinfo  {journal} {Linear Algebra and its Applications}\
  }\textbf {\bibinfo {volume} {212}},\ \bibinfo {pages} {413} (\bibinfo {year}
  {1994})}\BibitemShut {NoStop}%
\end{thebibliography}%

\end{document}